\documentclass[epsfig,12pt]{article}
\usepackage{makeidx}
\usepackage{amsmath}
\usepackage{amsfonts}
\usepackage{amssymb}
\usepackage{graphicx}

\input epsf.sty
\newtheorem{theorem}{Theorem}
\newtheorem{acknowledgement}[theorem]{Acknowledgement}

\makeatletter \@addtoreset{equation}{section}
\renewcommand{\theequation}{\thesection.\arabic{equation}}
\input{tcilatex}
\begin{document}

\title{\vspace{-2cm}%
\rightline{\mbox{\small {\bf LPHE-Preprint-December-2011
}}} \bigskip \bigskip \textbf{On Flavor Symmetry in Lattice Quantum
Chromodynamics}}
\author{El Hassan Saidi\thanks{%
E-mail: h-saidi@fsr.ac.ma}\bigskip \\
{\small 1}. {\small Lab Of High Energy Physics, Modeling and Simulations,
Faculty of Science, }\\
{\small University Mohammed V-Agdal, Av Ibn Battota, Rabat, Morocco}\\
{\small 2. Centre Of Physics and Mathematics, CPM- CNESTEN, Morocco}}
\maketitle

\begin{abstract}
Using a well established method to engineer non abelian symmetries in
superstring compactifications, we study the link between the point splitting
method of \emph{Creutz et al }of refs. \cite{A1,A2} for implementing flavor
symmetry in lattice QCD; and singularity theory in complex algebraic
geometry. We show amongst others that \emph{Creutz} flavors for naive
fermions are intimately related with toric singularities of a class of
complex Kahler manifolds that are explicitly built here. In the case of
naive fermions of QCD$_{2N}$, \emph{Creutz} flavors are shown to live at the
poles of real 2-spheres and carry quantum charges of the fundamental of $%
\left[ SU\left( 2\right) \right] ^{2N}$. We show moreover that the two \emph{%
Creutz} flavors in Karsten-Wilczek model, with Dirac operator in reciprocal
space of the form $i\mathbf{\gamma }_{1}\mathrm{F}_{1}+i\mathbf{\gamma }_{2}%
\mathrm{F}_{2}+$ $i\mathbf{\gamma }_{3}\mathrm{F}_{3}+$ $\frac{i}{\sin
\alpha }\mathbf{\gamma }_{^{4}}\mathrm{F}_{4}$, are related with the small
resolution of conifold singularity that live at $\sin \alpha =0$. Other
related features are also studied.\newline
\textbf{Key words}: {Naive and Karsten-Wilczek fermions, Point splitting
method, Toric geometry}.
\end{abstract}


\section{Introduction}

Recently M. Creutz and co-workers developed in refs \textrm{\cite{A1,A2}} a
method to implement flavor symmetry of quarks in lattice QCD by proposing a
nice interpretation to the different $\mathbf{\gamma }_{5}$- chiralities of
the zeros of the lattice Dirac operator as states of a flavor multiplet.\
This approach, known as \emph{the point-splitting method}, has been used for
various purposes; in particular to identify species in naive and minimally
doubled fermions as quark flavors with a non abelian symmetry; and also to
define proper flavored-mass terms to extract the index in the spectral flow 
\textrm{\cite{B1}; }see also\textrm{\  \cite{2B}-\cite{AB6} }for related
issues.\newline
On the other hand, one of the lessons learnt from the link between the gauge
theory of elementary particles and superstrings is the way to engineer non
abelian symmetries for both gauge invariance and flavors \textrm{\cite{C1,C2}%
}. The engineering of these symmetries has been shown to be a smart key to
approach the low energy limit of superstrings; especially in dealing with
Calabi-Yau compactifications of type II superstrings with branes wrapping
cycles \textrm{\cite{D0,D,1D,D1,D2}}. These compactifications involve
singular internal manifolds with local singularities leading remarkably to a
geometric engineering of the continuous symmetries that we see at low
energies \textrm{\cite{D3,D4,D5}}. \newline
The principal aim of this paper is work out explicitly the link between the
point-splitting method of Creutz and collaborators; and the geometric
engineering of symmetries by using singularity theory of complex geometry.
For concreteness, we will focuss on specific models of lattice QCD namely
the Karsten-Wilczek fermions and the naive ones; but our construction is
general and applies as well for other fermions. Among our results, we
mention too particularly the two following things. \newline
$1)$ the zero modes of the Dirac operator $D_{{\small naive}}$ for naive
fermions on 2d-dimension lattice $\mathcal{L}_{2d}$ are associated with 
\emph{toric singularities} \textrm{\cite{D0,E2,E3,E1} }of some Kahler
manifolds $\mathcal{K}_{d}$ to be constructed explicitly in section 3. The
unit cell $\mathcal{C}$ in the reciprocal lattice $\mathcal{\tilde{L}}_{2d}$
turn out to be exactly the real base $\Delta _{d}$ of the toric graph of the
corresponding complex Kahler manifold $\mathcal{K}_{d}$. To make an idea on
this strange link, recall that the expression of $D_{{\small naive}}$ reads
in the reciprocal space like 
\begin{equation}
D_{{\small naive}}=\dsum \limits_{l=1}^{2d}i\mathbf{\gamma }_{l}\sin p_{l}
\label{1}
\end{equation}%
with $\mathbf{\gamma }_{l}$ the hermitian $2^{d}\times 2^{d}$ gamma matrices
satisfying the Clifford algebra in 2d-dimensions. The zeros of this periodic
matrix operator, which are located at $\sin p_{l}=0$, that is at $p_{l}=0$
and $p_{l}=$ $\pi $ $\func{mod}2\pi $; may be remarkably thought of as the
north and south poles of a real 2-sphere $\mathbb{S}^{2}$; a property that
let understand that the Creutz flavors has much to do with the local patches
of $\mathbb{S}^{2}$; see details given in section 3 and eqs(\ref{st}-\ref{ts}%
) of appendix for further explicit relations. We will see\ throughout this
study that $D_{{\small naive}}$ can be also viewed as the antihermitian part
of the complex matrix operator%
\begin{equation}
\mathcal{D}=\dsum \limits_{l=1}^{2d}\frac{z_{l}}{\zeta _{l}}\mathbf{\gamma }%
_{l}  \label{2}
\end{equation}%
where $\left( z_{l},\zeta _{l}\right) \sim \left( \lambda z_{l},\lambda
\zeta _{l}\right) $, with arbitrary non zero $\lambda $, are homogeneous
complex coordinates parameterizing the complex projective line $CP^{1}\sim 
\mathbb{S}^{2}$. This manifold has two toric singularities effectively
located at the north and south poles of $\mathbb{S}^{2}$ and given, in
spherical coordinates $\left( x,y,z\right) =$ $\left( \sin \theta \cos
\varphi ,\sin \theta \sin \varphi ,\cos \varphi \right) $, by the solutions
of $\sin \theta =0$.

\  \  \  \newline
$2)$ The two zero modes of the Dirac for 4-dimensional Karsten-Wilczek
fermions have a different geometric interpretation with respect to those of
the naive fermions. In the Karsten-Wilczek case as described in \textrm{\cite%
{F1,F2,A1,A2}}, we find that the zeros are intimately related with the \emph{%
small resolution of the conifold singularity} in complex 3-dimension Kahler
manifolds \textrm{\cite{G1,G2,G3,G4}}.\ To exhibit rapidly this link, recall
that the Dirac of Karsten-Wilczek fermions reads in reciprocal space like 
\begin{equation}
D_{{\small KW}}=\dsum \limits_{l=1}^{3}i\mathbf{\gamma}_{l}\sin p_{l}+%
\mathbf{\gamma}_{4}\frac{i}{\sin \alpha}\left( \left( 1-\cos \alpha \right)
+\sum_{l=1}^{4}\left( 1-\cos p_{l}\right) \right)  \label{3}
\end{equation}
Notice that the $\gamma_{l}$- coefficients for the first three terms are
exactly as for naive fermions in 4d; and so capture the same geometrical
property as for (\ref{1}). The $\gamma_{4}$- coefficient however has a
different structure; it depends on two special things: $\left( i\right) $ it
is given by the sum over the terms $\left( 1-\cos p_{l}\right) $ that have
an interpretation in terms of the stereographic projection of the real
2-sphere; and $\left( ii\right) $ it has an extra real and free\textrm{%
\footnote{%
the original Karsten-Wilczek action corresponds to $\alpha=\frac{\pi}{2}$}}
parameter $\alpha$ showing that $D_{{\small KW}}$ is in fact an operator
flow with spectral parameter $\alpha$ encoding a remarkable singularity for 
\begin{equation}
\sin \alpha=0
\end{equation}
which, according to the expansions (\ref{p}-\ref{m}), it may be also
interpreted as the mass of a non relativistic mode living near the Dirac
points. We show in this study that for $\alpha=0$ $\func{mod}\pi$, this
singularity is exactly similar to the singularity of the conifold $T^{\ast}%
\mathbb{S}^{3}$; and the values $\alpha \neq0$ corresponds precisely to the
small resolution of the conic singularity at $\alpha=0$.\ 

\  \  \newline
The presentation is as follows: In section 2, we review briefly the main
lines of the point-splitting method of \emph{Creutz} for Karsten-Wilczek
(KW) fermions of lattice QCD$_{4}$; a similar construction is valid for
naive fermions (NF). In section 3, we study in details the link between the
zero modes of the Dirac operator of naive fermions of lattice QCD$_{2}$ and
toric singularities; the extension to higher dimensions is straightforward
and so omitted. In section 4, we give some useful tools on singularity
theory in complex geometry and work out the link with the Creutz point
splitting method. In section 5, we study the relation between the zero modes
of the Dirac operator of Karsten-Wilczek fermions and the small resolution
of conifold.\ In section 6, we give a conclusion and comments. In the
appendix, we recall some useful relations on stereographic projection of
real 2-sphere and its Kahler structure.

\section{Point-splitting method of Creutz}

In this section, we describe the main lines of the point splitting method of 
\emph{Creutz} \textrm{\cite{A1} }for the case of naive fermions and for the
Karsten-Wilczek ones on 4-dimensional lattice with typical Dirac operator in
reciprocal space as 
\begin{equation}
D_{lattice}=\dsum \limits_{l=1}^{4}i\mathbf{\gamma}_{l}\mathrm{F}_{l}
\label{f}
\end{equation}
For the naive fermions, the coefficients $\mathrm{F}_{l}=\frac{1}{2i}%
\left
\{ D_{lattice},\mathbf{\gamma}_{l}\right \} $ of the gamma matrices
are all of same nature; and are given by $\mathrm{F}_{l}=\sin p_{l}$ with%
\begin{equation}
p_{l}=\frac{k_{l}}{a}  \label{kp}
\end{equation}
For Karsten-Wilczek fermions, the $\mathrm{F}_{l}$'s are as in eq(\ref{3});
so the difference between $D_{lattice}^{\left( NF\right) }\equiv D_{NF}$ and 
$D_{lattice}^{\left( KW\right) }\equiv D_{KW}$ concerns the expression of
the component 
\begin{equation}
\mathrm{F}_{4}=\frac{1}{2i}\left \{ D_{{\small KW}},\mathbf{\gamma}%
_{4}\right \} .
\end{equation}
In eqs(\ref{f}-\ref{kp}), the $p_{l}$ variables are the phases of the wave
functions of the particle along the hopping direction; the $k_{l}$'s are the
components of the wave vector $\mathbf{k}=\left(
k_{1},k_{2},k_{3},k_{4}\right) $ and the number $a$ is the spacing parameter
of real lattice. In this section, we will mainly focus on KW fermions
because of its richer structure; a similar and straightforward analysis is
valid for the naive ones.

\subsection{specific features of Karsten-Wilczek fermion}

The Karsten-Wilczek fermion is a particular 4- dimensional QCD model living
on a hypercubic lattice with a Dirac operator $D_{KW}$ having two fermionic
zero modes. In the reciprocal space,\ the lattice Karsten-Wilczek operator $%
D_{KW}$ reads as follows%
\begin{equation}
D_{{\small KW}}=\dsum \limits_{l=1}^{3}i\mathbf{\gamma}_{l}\sin p_{l}+%
\mathbf{\gamma}_{4}\frac{i}{\sin \alpha}\left( \cos
\alpha+3-\sum_{l=1}^{4}\cos p_{l}\right) ,  \label{ope}
\end{equation}
where, for simplicity, interactions with link fields have been dropped out.

\emph{Special properties of }$D_{{\small KW}}$\newline
The operator $D_{{\small KW}}$ has some remarkable features; in particular
the 3 following ones relevant for our study: $\left( i\right) $ it depends
on an extra real free parameter $\alpha$ whose geometric interpretation will
be studied in details later on. $\left( ii\right) $ It is an anti-hermitian
operator that follows from the Fourier transform of the tight binding
hamiltonian of Karsten-Wilczek fermions \textrm{\cite{A2,H1,H2}. }This
means, on one hand, that $D_{{\small KW}}$ can be imagined as 
\begin{equation}
D_{{\small KW}}=\frac{1}{2}\left( D-D^{+}\right)
\end{equation}
and, on the other hand, it has a hermitian companion $D^{sym}=\frac{1}{2}%
\left( D+D^{+}\right) $ that is expected to play some role in the geometric
interpretation of the zeros. $\left( iii\right) $ Because of the gamma
matrices, $D_{{\small KW}}$ is $4\times4$ matrix operator that acts on 4-
component spinorial wave functions $\psi \left( \mathbf{k}\right) $
depending on the wave vectors $\mathbf{k}$ of the hopping particles,%
\begin{equation}
\psi \left( \mathbf{k}\right) =\left( 
\begin{array}{c}
\phi_{i}\left( \mathbf{k}\right) \\ 
\bar{\chi}_{i}\left( \mathbf{k}\right)%
\end{array}
\right)
\end{equation}
with $\phi_{i}$ and $\bar{\chi}_{i}$ standing respectively for the left $%
\psi_{L}=\frac{1}{2}\left( 1+\gamma_{5}\right) \psi$ and the right $%
\psi_{R}= $ $\frac{1}{2}\left( 1-\gamma_{5}\right) \psi$ handed of $\psi$.
The latters are 2 component Weyl spinors of $SO\left( 4\right) $,%
\begin{equation}
\begin{tabular}{lll}
$\phi_{i}=\left( 
\begin{array}{c}
\phi_{1}\left( \mathbf{k}\right) \\ 
\phi_{2}\left( \mathbf{k}\right)%
\end{array}
\right) $ & , & $\bar{\chi}_{\bar{\imath}}=\left( 
\begin{array}{c}
\bar{\chi}_{1}\left( \mathbf{k}\right) \\ 
\bar{\chi}_{2}\left( \mathbf{k}\right)%
\end{array}
\right) $%
\end{tabular}%
\end{equation}
having opposite chiralities; $\gamma_{5}\psi_{L}=\psi_{L}$ and $%
\gamma_{5}\psi_{R}=-\psi_{R}$. The operator $D_{{\small KW}}$ has two zero
modes which we denote as $\mathbf{P}^{\pm}=a\mathbf{K}^{\pm}$; they are
located in the reciprocal space at 
\begin{equation}
\begin{tabular}{llll}
$\mathbf{P}^{+}=\left( 0,0,0,+\alpha \right) $ & , & $\mathbf{P}^{-}=\left(
0,0,0,-\alpha \right) $ & .%
\end{tabular}
\label{pm}
\end{equation}
The propagator $\left \langle \bar{\psi}\left( p\right) \psi \left( p\right)
\right \rangle =\frac{D_{{\small KW}}}{D_{{\small KW}}^{2}}$ has \emph{%
coupled simple poles}; by setting $p_{1}=$ $p_{2}=$ $p_{3}=0$ for simplicity
and leaving $p_{4}$ free, one can show that this propagator reads as 
\begin{equation}
\left \langle \bar{\psi}\left( p\right) \psi \left( p\right) \right \rangle
|_{_{p_{1}=p_{2}=p_{3}=0}}=\frac{-i\gamma_{4}\cot^{2}\frac{\alpha}{2}}{%
\left( 1-\frac{\sin \frac{p}{2}}{\sin \frac{\alpha}{2}}\right) \left( 1+%
\frac {\sin \frac{p}{2}}{\sin \frac{\alpha}{2}}\right) }  \label{tf}
\end{equation}
Observe by the way that in the limit $\alpha=0$, the two zeros of eq(\ref{pm}%
) collide at the origin $\left( 0,0,0,0\right) $ of the reciprocal space.
This leads to a double pole and so to a symmetry enhancement to be studied
in section 5.

\emph{Expansions of }$D_{{\small KW}}$\emph{\ near the zero modes}\newline
To get the expressions of the Dirac operator near its two zero modes $P_{\mu
}^{\pm}$, we first set $p_{\mu}=P_{\mu}^{\pm}+q_{\mu}$ with $q_{\mu}$ small
fluctuations around the zeros; then expand $D_{KW}$ in series of $q_{\mu}$
to end with two $4\times4$ matrix operators that we denote like 
\begin{equation}
\begin{tabular}{lll}
$D_{+}=D\left( p_{\mu}-P_{\mu}^{+}\right) $ & , & $D_{-}=D\left( p_{\mu
}-P_{\mu}^{-}\right) $%
\end{tabular}%
\end{equation}
The value of these operators at first orders in $q_{\mu}$ gives the usual
Dirac operator in continuum; but with two different kinds of gamma matrix
representations $\mathbf{\gamma}_{\mu}$ and $\mathbf{\gamma}_{\mu}^{\prime}$%
. For $D_{+}$, we have at the two leading orders in $q_{\mu}$ 
\begin{equation}
D_{+}=\dsum \limits_{\mu=1}^{4}i\mathbf{\gamma}_{\mu}q_{\mu}+i\mathbf{\gamma}%
_{4}\left( \dsum \limits_{\mu=1}^{4}\frac{\left( q_{\mu}\right) ^{2}}{2\sin
\alpha}\right) +O\left( q^{3}\right)  \label{p}
\end{equation}
where the first term is precisely the Dirac operator of a free particle and
the second term could be interpreted as a non relativistic massive mode with
mass depending on $\alpha$. Similarly, we find for $D_{-}$,%
\begin{equation}
D_{-}=\dsum \limits_{\mu=1}^{4}i\mathbf{\gamma}_{\mu}^{\prime}q_{\mu}+i%
\mathbf{\gamma}_{4}^{\prime}\left( \dsum \limits_{\mu=1}^{4}\frac{\left(
q_{\mu}\right) ^{2}}{2\sin \alpha}\right) +O\left( q^{3}\right)  \label{m}
\end{equation}
but now with different matrices $\mathbf{\gamma}_{\mu}^{\prime}$ that are
related to the previous $\mathbf{\gamma}_{\mu}$'s as%
\begin{equation}
\begin{tabular}{lllll}
$\mathbf{\gamma}_{1}^{\prime}=\mathbf{\gamma}_{1}$ & , & $\mathbf{\gamma}%
_{2}^{\prime}=\mathbf{\gamma}_{2}$ & , & $\mathbf{\gamma}_{3}^{\prime }=%
\mathbf{\gamma}_{3}$%
\end{tabular}%
\end{equation}
and%
\begin{equation}
\begin{tabular}{llll}
$\mathbf{\gamma}_{4}^{\prime}=-\mathbf{\gamma}_{4}$ & , & $\mathbf{\gamma}%
_{5}^{\prime}=-\mathbf{\gamma}_{5}$ & 
\end{tabular}%
\end{equation}
showing that the two zeros (\ref{pm}) are not equivalent as they have
opposite $\mathbf{\gamma}_{5}$-chiralities. Notice that the relations
between $\mathbf{\gamma}_{\mu}^{\prime}$ and $\mathbf{\gamma}_{\mu}$ can be
rewritten into a condensed form like 
\begin{equation}
\mathbf{\gamma}_{\mu}^{\prime}=\Gamma^{+}\mathbf{\gamma}_{\mu}\Gamma
\end{equation}
where $\Gamma$ is a similarity transformation given by $\Gamma=i\mathbf{%
\gamma }_{4}\mathbf{\gamma}_{5}$ and preserve the Clifford algebra $\left \{
\gamma_{\mu},\gamma_{\nu}\right \} =\left \{ \gamma_{\mu}^{\prime},\gamma
_{\nu}^{\prime}\right \} =2\delta_{\mu \nu}$. The last feature can be
explicitly checked by using the following representation of the gamma
matrices%
\begin{equation}
\begin{tabular}{lll}
$\gamma^{k}{\small =}\left( 
\begin{array}{cc}
{\small 0} & {\small -i\sigma}^{k} \\ 
{\small i\sigma}^{k} & {\small 0}%
\end{array}
\right) $ & , & $\gamma^{4}{\small =}\left( 
\begin{array}{cc}
{\small 0} & {\small I} \\ 
{\small I} & {\small 0}%
\end{array}
\right) \ $%
\end{tabular}%
\end{equation}
with ${\small \sigma}^{k}$ standing for the usual $2\times2$ Pauli matrices
and 
\begin{equation}
\begin{tabular}{lll}
$\gamma^{5}{\small =}\left( 
\begin{array}{cc}
{\small I} & {\small 0} \\ 
{\small 0} & {\small -I}%
\end{array}
\right) $ & , & $\Gamma=\left( 
\begin{array}{cc}
{\small 0} & -i{\small I} \\ 
i{\small I} & {\small 0}%
\end{array}
\right) $%
\end{tabular}%
\end{equation}
from which we learn amongst others that $\Gamma^{+}=\Gamma$ and $\Gamma
^{+}\Gamma=I_{4\times4}$.

\subsection{the point splitting method}

The point splitting method of Creutz \textrm{\cite{A1,A2}} identifies the
two \emph{inequivalent} species of the KW fermions that are associated with
the operators $D_{+}$ and $D_{-}$ as two independent flavors denoted as $u$
and $d$. Each flavor field is defined so that the associated fermion
propagator, namely $\left \langle \bar{u}\left( p\right) u\left( p\right)
\right \rangle $ and $\left \langle \bar{d}\left( p\right) d\left( p\right)
\right \rangle $, includes only a single and simple pole. Recall that the
propagator of KW fermions has coupled poles (\ref{tf}) having the typical
form%
\begin{equation}
\left \langle \bar{\psi}\left( p\right) \psi \left( p\right) \right \rangle
|_{_{p_{1}=p_{2}=p_{3}=0}}=\frac{2C}{\left( Z-1\right) \left( Z+1\right) }
\end{equation}
with some number $C$. These coupled simple poles at $Z=\pm1$ can be split in
terms of two isolated simple poles by using the relation 
\begin{equation}
\frac{2C}{\left( Z-1\right) \left( Z+1\right) }=\frac{C}{Z-1}-\frac {C}{Z+1}
\end{equation}
Following \textrm{\cite{A1,A2}}, the point splitting method to get the
flavor $u\left( \mathbf{p}\right) $ living in the neighborhood of $%
P_{\mu}^{+}$ can be done by multiplying the wave function $\psi \left( 
\mathbf{p}\right) $ by a factor that removes the other pole at $P_{\mu}^{-}$%
. The same procedure can be done for the flavor $d\left( \mathbf{p}\right) $%
. This procedure leads to the fields 
\begin{equation}
\begin{tabular}{lll}
$u\left( \mathbf{p}-\alpha \mathbf{e}_{4}\right) $ & $=$ & $\frac{1}{2}%
\left( 1+\frac{\sin p_{4}}{\sin \alpha}\right) \psi \left( \mathbf{p}\right) 
$ \\ 
&  &  \\ 
$d\left( \mathbf{p}+\alpha \mathbf{e}_{4}\right) $ & $=$ & $\frac{1}{2}%
\left( 1-\frac{\sin p_{4}}{\sin \alpha}\right) \Gamma \psi \left( \mathbf{p}%
\right) $%
\end{tabular}
\label{ud}
\end{equation}
with $\Gamma=i\mathbf{\gamma}_{4}\mathbf{\gamma}_{5}$; and are thought of as
the components of an SU$\left( 2\right) $ flavor doublet,%
\begin{equation}
\Psi \left( \mathbf{p}\right) =\left( 
\begin{array}{c}
u\left( \mathbf{p}-\alpha \mathbf{e}_{4}\right) \\ 
d\left( \mathbf{p}+\alpha \mathbf{e}_{4}\right)%
\end{array}
\right)  \label{d}
\end{equation}
Here $\mathbf{e}_{1}$, $\mathbf{e}_{2}$, $\mathbf{e}_{3}$, $\mathbf{e}_{4}$
stand for the basis vectors in the 4-dimensional reciprocal space. In the
doubler representation $\Psi$, the usual chiral matrix $\gamma_{5}$ acting
on $\psi$ gets promoted to the tensor product 
\begin{equation}
\begin{tabular}{lll}
$\gamma_{5}\otimes \tau^{3}$ & , & $\tau^{3}=\left( 
\begin{array}{cc}
1 & 0 \\ 
0 & -1%
\end{array}
\right) $%
\end{tabular}%
\end{equation}
together with 
\begin{equation}
\begin{tabular}{lll}
$\gamma_{5}\psi$ & $\rightarrow$ & $\left( 
\begin{array}{cc}
\gamma_{5} & 0 \\ 
0 & -\gamma_{5}%
\end{array}
\right) \Psi$%
\end{tabular}%
\end{equation}
where now $\Psi$ is as in (\ref{d}). This point splitting method allows to
construct as well mass terms (flavored-mass terms) that assign different
masses to the two species. This is done by promoting the 4$\times$4 identity
matrix $I_{4}$ on $\psi$ as follows%
\begin{equation}
\begin{tabular}{lll}
$I_{4}.\psi$ & $\rightarrow$ & $\left( I_{4}\otimes \tau^{3}\right)
\Psi=\left( 
\begin{array}{cc}
I_{4} & 0 \\ 
0 & -I_{4}%
\end{array}
\right) \Psi$%
\end{tabular}%
\end{equation}
so that%
\begin{equation}
\begin{tabular}{lll}
$\bar{\psi}I_{4}\psi$ & $\rightarrow$ & $\bar{\Psi}\left( I_{4}\otimes
\tau^{3}\right) \Psi$%
\end{tabular}%
\end{equation}
which up on substituting (\ref{d}) leads to, on one hand, to 
\begin{equation}
\bar{\Psi}\left( I_{4}\otimes \tau^{3}\right) \Psi=\bar{u}u-\bar{d}d
\end{equation}
and by help of (\ref{ud}) to 
\begin{equation}
\bar{u}u-\bar{d}d=\frac{\sin p_{4}}{\sin \alpha}\bar{\psi}\psi,\qquad \bar {u%
}u+\bar{d}d=\bar{\psi}\psi
\end{equation}
A similar analysis is valid for naive fermions; for more details see below
and \textrm{\cite{A2}}. In what follows, we study the link between the point
splitting method of Creutz and a class of singular complex Kahler manifolds.
We first consider the case of naive fermions in 2-dimensional lattice as a
matter to illustrate the idea and also to introduce some useful tools; then
we turn to the KW model.

\section{Naive fermions and toric singularities}

In this section we show that the point splitting method for naive fermions
(NF) is intimately associated with toric singularities of a particular class
of toric manifolds. For the example of 2-dimensional naive fermions on which
we focuss below, the corresponding toric manifold is precisely given by the
complex projective surface 
\begin{equation}
\mathcal{S}=CP^{1}\times CP^{1}
\end{equation}%
made of two $CP^{1}$ copies. Recall that the complex projective line $CP^{1}$
is isomorphic the usual real 2-sphere $\mathbb{S}^{2}$ and therefore $%
\mathcal{S}$ is given by the real 4-dimensional compact manifold $\mathbb{S}%
^{2}\times \mathbb{S}^{2}$ \textrm{\cite{I1}}. We show also that the first
Brillouin zone modded by $\mathbb{Z}_{2}$ symmetry of the Dirac operator is
nothing but the toric graph of $CP^{1}\times CP^{1}$.

\subsection{the naive $D_{NF}$ operator and toric manifolds}

To exhibit the link between the zeros of the Dirac operator of naive
fermions and toric singularities, let us focus on the simple case of a
2-dimensional lattice with Dirac operator $D_{NF}$ in reciprocal space given
by the following field matrix, 
\begin{equation}
D_{NF}=i\gamma_{1}\sin p_{1}+i\gamma_{2}\sin p_{2}  \label{dn}
\end{equation}
with the phases $\left( p_{1},p_{2}\right) $ related to the wave vector
components $\left( k_{1},k_{2}\right) $ as in eq(\ref{kp}); that is $%
p_{1}=ak_{1},$ $p_{2}=ak_{2}$. The extension to higher dimensions is
straightforward.

\subsubsection{explicit features of $D_{NF}$ in QCD$_{2}$}

\begin{figure}[ptbh]
\begin{center}
\hspace{0cm} \includegraphics[width=8cm]{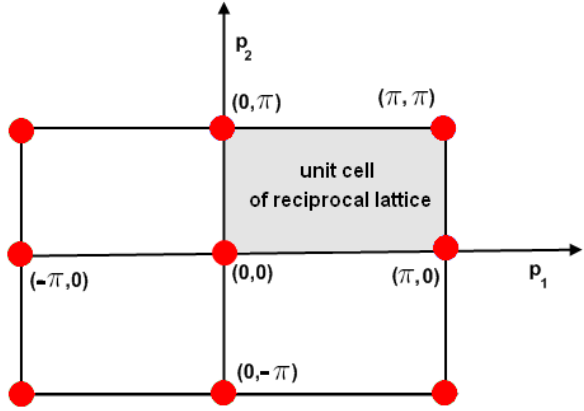}
\end{center}
\par
\vspace{-0.5cm}
\caption{fundamental domain in reciprocal space. $\mathbb{Z}_{2}$ symmetry
splits this domain into 4 unit cells. The vertices describe the zeros of $%
D_{naive}$ of QCD$_{2}$ fermions; they are also the fix points of $\mathbb{Z}%
_{2}$. }
\label{CU}
\end{figure}

First notice that to make contact between the zeros of $D_{NF}$ and the
usual singularities of complex algebraic geometry, we need to identify a
complex operator $\mathcal{D}$ that is associated with $D_{NF}$ and
capturing the same physical features as those carried by (\ref{dn}). In
particular, it has to be valued in the Clifford algebra in same manner as $%
D_{NF}$; 
\begin{equation}
\mathcal{D}=\gamma_{1}\mathcal{F}_{1}\left( z_{1},z_{2}\right) +\gamma _{2}%
\mathcal{F}_{2}\left( z_{1},z_{2}\right)  \label{ff}
\end{equation}
and has to live on same complex surface parameterized by local coordinates $%
\left( z_{1},z_{2}\right) $ with $\mathcal{F}_{1}\left( z_{1},z_{2}\right) $
and $\mathcal{F}_{2}\left( z_{1},z_{2}\right) $ having the same zeros as $%
D_{NF}.$ \newline
To build this complex operator, we start by analyzing some useful properties
of $D_{NF}$ that are behind the point splitting method. These features are
of two kinds: those manifestly exhibited on (\ref{dn}); and others ones
implicit. Here, we consider the manifest properties of $D_{NF}$; the
implicit ones will be given in next subsection. \newline
From eq(\ref{dn}), we learn that $D_{NF}$ has 3 remarkable and manifest
properties: \newline
$\left( 1\right) $ it is an anti-hermitian operator%
\begin{equation}
D_{NF}^{+}=-D_{NF}  \label{an}
\end{equation}
and so it can be put into the form 
\begin{equation}
D_{NF}=\frac{1}{2}\left( D-D^{+}\right)
\end{equation}
with 
\begin{equation}
D=\gamma_{1}e^{ip_{1}}+\gamma_{2}e^{ip_{2}}.  \label{e}
\end{equation}
This feature means also that $D_{NF}$ has a partner given by 
\begin{equation}
D^{\left( sym\right) }=\frac{1}{2}\left( D+D^{+}\right)
\end{equation}
that turns out to be helpful in looking for a geometric interpretation of
the origin of the splitting method of Creutz. \newline
$\left( 2\right) $ $D_{NF}$ is generally non linear in $\left(
p_{1},p_{2}\right) $ as it depends on the $\sin p_{l}$'s; but its expansion
near the zeros 
\begin{equation*}
p_{l}=0,\text{ }\pi
\end{equation*}
has a linear behavior precisely given by the Dirac operator in continuum
namely 
\begin{equation}
i\gamma_{1}p_{1}+i\gamma_{2}p_{2}.
\end{equation}
This property is also required by lattice QCD in the continuous limit.%
\newline
$\left( 3\right) $ $D_{NF}$ is periodic in both $p_{1}$ and $p_{2}$
variables; a property that allows to restrict the analysis in the reciprocal
space to the fundamental surface of the 2-dimensional momentum space 
\begin{equation}
\begin{tabular}{llll}
$-\pi \leq p_{l}\leq \pi$ & , & $-b\pi \leq k_{l}\leq b\pi$ & 
\end{tabular}%
\end{equation}
with $b=\frac{1}{a}$ the spacing parameter of the reciprocal lattice; see
also fig \ref{CU}. 

\subsubsection{implicit features of $D_{NF}$}

\begin{figure}[ptbh]
\begin{center}
\hspace{0cm} \includegraphics[width=8cm]{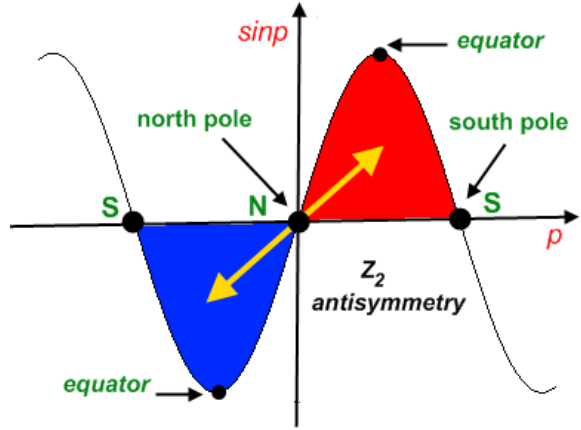}
\end{center}
\par
\vspace{-1cm}
\caption{zeros of $\sin p$ as north and south poles of a real 2-sphere;
parallel circle at the equatorial plane shrinks at the poles N and S.}
\label{S}
\end{figure}

The $D_{NF}$ given by eq(\ref{dn}) has also non manifest features; one of
them is that it behaves as an odd operator under the $\mathbb{Z}_{2}$
transformation in the momentum space 
\begin{equation}
\left( p_{1},p_{2}\right) \rightarrow \left( -p_{1},-p_{2}\right) ,\qquad
\left( k_{1},k_{2}\right) \rightarrow \left( -k_{1},-k_{2}\right)  \label{Z2}
\end{equation}
This property allows to restrict the analysis on $D_{NF}$ further to 
\begin{equation}
\begin{tabular}{lll}
$0\leq p_{l}\leq \pi$ & , & $0\leq k_{l}\leq b\pi$%
\end{tabular}%
\end{equation}
and to which we refer to as unit cell; see fig \ref{CU}. Later on, we will
show that this unit cell is precisely the toric graph of the complex
projective surface $CP^{1}\times CP^{1}$ with shrinking 1-cycle on edges and
2-cycle at the vertices. This feature is related to the fact the two zeros
of the $\sin p$'s of (\ref{dn}) can be interpreted as the north and south
poles of a real 2-sphere with coordinates 
\begin{equation}
\mathrm{x}=\sin p\cos \varphi,\qquad \mathrm{y}=\sin p\sin \varphi,\qquad 
\mathrm{z}=\cos p
\end{equation}
with $\varphi$ generating the circle $\mathbb{S}_{\varphi}^{1}$ 
\begin{equation}
\mathrm{x}^{2}+\mathrm{y}^{2}=\rho^{2},\qquad \rho=\sin p  \label{313}
\end{equation}
For $p=0$, this circle $\mathbb{S}_{\varphi}^{1}$ shrinks to the north pole
N located at 
\begin{equation}
\left( \mathrm{x},\mathrm{y},\mathrm{z}\right) =\left( 0,0,1\right)
\end{equation}
and for $p=\pi$ it shrinks to the south pole S at 
\begin{equation}
\left( \mathrm{x},\mathrm{y},\mathrm{z}\right) =\left( 0,0,-1\right)
\end{equation}
The variation of the radius $\rho=\sin p$ of the circle $\mathbb{S}_{\varphi
}^{1}$ in terms of the phase of the particle is depicted on fig \ref{S}. 
Another implicit feature of $D_{NF}$ concerns the linking of the point
splitting method to singularity theory in complex geometry. The complex
expression (\ref{e}) leading to $D_{NF}$ looks a very particular quantity
since the natural complex extension of $D_{NF}$ would be like 
\begin{equation}
\boldsymbol{D}=z_{1}\gamma_{1}+z_{2}\gamma_{2},  \label{dz}
\end{equation}
which, by requiring $\left \vert z_{1}\right \vert =\left \vert
z_{2}\right
\vert =1$, one recovers eq(\ref{e}). However, this complex
extension cannot play the role of $\mathcal{D}$ given by eq(\ref{ff}). The
expression (\ref{dz}) destroys a main property of naive fermions since one
looses a basic information about the number of zeros of the operator $D_{NF}$%
. This means that eq(\ref{dz}) is not the exact complex extension of $D_{NF}$%
. Contrary to (\ref{dn}) which has 4 zeros, the operator $\boldsymbol{D}$
has only one zero located at the origin%
\begin{equation}
\left( z_{1},z_{2}\right) =\left( 0,0\right)
\end{equation}
of the complex surface with local coordinates $\left( z_{1},z_{2}\right) $.
A careful inspection shows that the exact complex extension of $D_{NF}$
encoding all data on the zeros of $D_{NF}$ should be as follows:%
\begin{equation}
\mathcal{D}=\frac{z_{1}}{\zeta_{1}}\gamma_{1}+\frac{z_{2}}{\zeta_{2}}%
\gamma_{2}  \label{D}
\end{equation}
where $\left( z_{i},\zeta_{i}\right) \in \mathbb{C}-\left \{ \left(
0,0\right) \right \} $.\textrm{\ }Notice that this complex operator\textrm{\ 
}$\mathcal{D}$ is invariant under the gauge symmetry\textrm{\ }%
\begin{equation}
\begin{tabular}{lll}
$\left( z_{1},\zeta_{1}\right) $ & $\rightarrow$ & $\left( \lambda
z_{1},\lambda \zeta_{1}\right) $ \\ 
$\left( z_{2},\zeta_{2}\right) $ & $\rightarrow$ & $\left( \mu z_{2},\mu
\zeta_{2}\right) $%
\end{tabular}%
\end{equation}
with $\lambda$ and $\mu$ two non zero complex number generating the $\mathbb{%
C}^{\ast}\times \mathbb{C}^{\ast}$ symmetry group. From this view, the
operator $\boldsymbol{D}=z_{1}\gamma_{1}+z_{2}\gamma_{2}$ appears as a
particular case of (\ref{D})\ and is recovered by fixing $\zeta_{1}$ and $%
\zeta_{2}$ as follows:%
\begin{equation}
\begin{tabular}{lll}
$\zeta_{1}=1$ & , & $\zeta_{2}=1$%
\end{tabular}
\label{cd}
\end{equation}
The others zeros of the naive fermions are given by working out the full set
of solutions of $\mathcal{D}=0$; this is done below.

\subsection{Zeros of complexified naive $\mathcal{D}$ and point splitting
method}

We first study the zeros of $\mathcal{D}$ by using the homogeneous complex
coordinates $\left( z_{1},\zeta_{1},z_{2},\zeta_{2}\right) $ of the complex
surface $\mathcal{S}$; then we give the link between the patches of $%
\mathcal{S}$ and the point splitting method.

\subsubsection{the 4 zeros of\emph{\ }$\mathcal{D}$}

The conditions (\ref{cd}) on the complex variables $\zeta_{1}$ and $%
\zeta_{2} $ are well known relations in complex projective geometry; they
correspond precisely to a particular gauge fixing of the $\mathbb{C}%
^{\ast}\times \mathbb{C}^{\ast}$ gauge symmetry of $\mathcal{D}$ 
\begin{equation}
\begin{tabular}{lll}
$\left( z_{1},\zeta_{1}\right) $ & $\rightarrow$ & $\left( z_{1}^{\prime
},\zeta_{1}^{\prime}\right) =\left( \lambda z_{1},\lambda \zeta_{1}\right) $
\\ 
$\left( z_{2},\zeta_{2}\right) $ & $\rightarrow$ & $\left( z_{2}^{\prime
},\zeta_{2}^{\prime}\right) =\left( \mu z_{2},\mu \zeta_{2}\right) $%
\end{tabular}
\label{cp}
\end{equation}
Being arbitrary non zero complex numbers, one may use this arbitrariness to
make diverse particular choices of $\lambda$ and $\mu$. One of these choices
is given by the example%
\begin{equation}
\begin{tabular}{lll}
$\lambda \zeta_{1}=1$ & $,$ & $\mu \zeta_{2}=1$%
\end{tabular}
\label{ch}
\end{equation}
leading to the following local coordinate patch of the surface surface $%
\mathcal{S}$%
\begin{equation}
\begin{tabular}{lll}
$\left( z_{1}^{\prime},\zeta_{1}^{\prime}\right) =\left( z,1\right) $ & , & $%
z=\frac{z_{1}}{\zeta_{1}}$ \\ 
$\left( z_{2}^{\prime},\zeta_{2}^{\prime}\right) =\left( w,1\right) $ & , & $%
w=\frac{z_{2}}{\zeta_{2}}$%
\end{tabular}%
\end{equation}
These relations teach us that the complex variables $\left( z_{1},\zeta
_{1},z_{2},\zeta_{2}\right) $ are \emph{homogeneous} coordinates that
parameterize the complex projective surface%
\begin{equation}
\mathcal{S}=CP^{1}\times CP^{1},  \label{s}
\end{equation}
embedded in the complex space $\mathbb{C}^{4}$. They tell us moreover that $%
\left( z_{1},\zeta_{1}\right) $ and $\left( z_{2},\zeta_{2}\right) $ are
respectively the complex coordinates parameterizing the complex projective
curves $CP^{1}$ of the complex surface (\ref{s}). \newline
Notice that, along with (\ref{ch}), there are also 3 other possible and
independent choices of the gauge parameters $\lambda$ and $\mu$; they lead
to 3 other local patches of the complex surface $\mathcal{S}$; they will be
given later on. Notice also that in the language of real geometry, the
manifold $\mathcal{S}$ can be thought of as given by the 4-real dimensional
compact space $\mathcal{S}$ $=$ \ $\mathbb{S}^{2}\times \mathbb{S}^{2}$
which can be viewed as 
\begin{equation}
\mathcal{S}\text{ }=\text{ \ }\mathbb{S}_{f}^{2}\times \mathbb{S}_{b}^{2}
\end{equation}
describing the fibration of a real 2-sphere $\mathbb{S}_{f}^{2}$ fibered on
the base 2-sphere $\mathbb{S}_{b}^{2}$. \newline
The complex operator $\mathcal{D}$ has 4 manifest zeros located at 
\begin{equation}
\left( z_{1},\frac{1}{\zeta_{1}},z_{2},\frac{1}{\zeta_{2}}\right) =\left \{ 
\begin{array}{c}
\left( 0,1,0,1\right) \\ 
\left( 0,1,1,0\right) \\ 
\left( 1,0,0,1\right) \\ 
\left( 1,0,1,0\right)%
\end{array}
\right.  \label{zz}
\end{equation}
in agreement with (\ref{e})\textrm{.} These 4 zeros are precisely located at
the north $N$ and south $S$ poles of the spheres $\mathbb{S}_{f}^{2}\times 
\mathbb{S}_{b}^{2}$ namely 
\begin{equation}
\begin{tabular}{llll}
$\left( N_{f},N_{b}\right) ,$ & $\left( N_{f},S_{b}\right) ,$ & $\left(
S_{f},N_{b}\right) ,$ & $\left( S_{f},S_{b}\right) $%
\end{tabular}%
\end{equation}
where live as well toric singularities; for technical details see section 4.

\subsubsection{point splitting method}

Invariance of $\mathcal{D}$ under the gauge symmetry (\ref{cd}) allows to
make appropriate choices of local coordinate patches $\mathcal{U}$ of the
complex surface $\mathcal{S}=CP^{1}\times CP^{1}$ without affecting the
physical properties carried by $\mathcal{D}$. It happens that there are 4
main possible and independent local coordinate patches%
\begin{equation}
\begin{tabular}{llll}
$\mathcal{U}_{I},$ \  & $\mathcal{U}_{II},$ \  & $\mathcal{U}_{III},$ \  & $%
\mathcal{U}_{IV}$%
\end{tabular}%
\end{equation}
with the intersection property%
\begin{equation}
\begin{tabular}{lll}
$\mathcal{U}_{I}$ $\cap$ $\mathcal{U}_{II}$ $\  \subseteq$ $\  \mathbb{C}$ & $%
, $ \  \  \  \  \  & $\mathcal{U}_{III}$ $\cap$ $\mathcal{U}_{IV}$ $\  \subseteq$ 
$\  \mathbb{C}$%
\end{tabular}%
\end{equation}
These patches are also in one to one correspondence with the 4 possible
gauge fixing conditions of the $\mathbb{C}^{\ast}\times \mathbb{C}^{\ast}$
symmetry%
\begin{equation}
\begin{tabular}{lllll}
$\mathcal{U}_{I}$ & : & $\lambda \zeta_{1}=1$ & $,$ & $\mu \zeta_{2}=1$ \\ 
$\mathcal{U}_{II}$ & : & $\lambda \zeta_{1}=1$ & $,$ & $\mu z_{2}=1$ \\ 
$\mathcal{U}_{III}$ & : & $\lambda z_{1}=1$ & $,$ & $\mu \zeta_{2}=1$ \\ 
$\mathcal{U}_{IV}$ & : & $\lambda z_{1}=1$ & $,$ & $\mu z_{2}=1$%
\end{tabular}
\  \   \label{gc}
\end{equation}
For example, on the coordinate patch $\mathcal{U}_{I}$ of the complex
surface $\mathcal{S}=CP^{1}\times CP^{1}$, which is also equivalent to just
setting%
\begin{equation}
\mathcal{U}_{I}:\qquad \zeta_{1}=1,\qquad \zeta_{2}=1
\end{equation}
and thinking about $z_{1}$ and $z_{2}$ as free local complex coordinates,
the gauge symmetry $\mathbb{C}^{\ast}\times \mathbb{C}^{\ast}$ is completely
fixed; and the complexified Dirac operator $\mathcal{D}$ reduces to eq(\ref%
{dz}) with a zero at $\left( 0,1,0,1\right) $. This property shows that the
splitting method of the four zeros of (\ref{dn}) is intimately related with
the gauge fixing choices (\ref{gc}). In other words, there is a one to one
correspondence between the 4 Creutz flavors and the Dirac operator $\mathcal{%
D}\left( \mathcal{U}\right) $ on the 4 local patches of $\mathcal{S}=\mathbb{%
S}^{2}\times \mathbb{S}^{2}$ as given in the following table,

\begin{align*}
& 
\begin{tabular}{l|l|l|l}
patch of $\mathcal{S}$ \  & $\  \left( z_{1},\zeta_{1}\right) \in CP_{\left(
1\right) }^{1}$ \  & $\  \left( z_{2},\zeta_{2}\right) \in CP_{\left(
2\right) }^{1}$ \  & $\  \  \  \  \  \  \  \  \  \  \  \  \mathcal{D}$ \\ \hline
$\  \  \  \  \  \mathcal{U}_{I}$ & $\  \  \  \  \  \left( z_{1},1\right) $ & $\  \  \  \
\  \left( z_{2},1\right) $ & $\  \mathcal{D}_{I}=z_{1}\gamma
_{1}+z_{2}\gamma_{2}$ \\ 
&  &  &  \\ 
$\  \  \  \  \  \mathcal{U}_{II}$ & $\  \  \  \  \  \left( z_{1},1\right) $ & $\  \  \  \
\  \left( 1,\zeta_{2}\right) $ & $\  \mathcal{D}_{II}=z_{1}\gamma _{1}+\frac{1%
}{\zeta_{2}}\gamma_{2}$ \\ 
&  &  &  \\ 
$\  \  \  \  \  \mathcal{U}_{III}$ & $\  \  \  \  \  \left( 1,\zeta_{1}\right) $ & $\
\  \  \  \  \left( z_{2},1\right) $ & $\  \mathcal{D}_{III}=\frac{1}{\zeta _{1}}%
\gamma_{1}+z_{2}\gamma_{2}$ \\ 
&  &  &  \\ 
$\  \  \  \  \  \mathcal{U}_{IV}$ & $\  \  \  \  \  \left( 1,\zeta_{1}\right) $ & $\  \
\  \  \  \left( 1,\zeta_{2}\right) $ & $\  \mathcal{D}_{IV}=\frac{1}{\zeta_{1}}%
\gamma_{1}+\frac{1}{\zeta_{2}}\gamma_{2}$ \\ \hline
\end{tabular}
\\
&
\end{align*}
On each local patch $\mathcal{U}$ of the complex surface $\mathcal{S}$, the
complex operator $\mathcal{D}\left( \mathcal{U}\right) $ has a simple zero
where live a degenerate \emph{toric symmetry}. \newline
Below, we give details on this kind of singularities whose typical graphic
representation is depicted in figures \ref{TA} and \ref{TB}.

\section{Toric singularities and Creutz Flavors}

Here we show, by using explicit tools, that Creutz flavors in naive fermions
live exactly at the toric singularities of the complex surface $CP^{1}\times
CP^{1}$; the extension to higher dimensions follows the same rule. To that
purpose, we start by recalling some useful tools on toric symmetry and toric
singularities; then we make the link with Creutz flavors.

\subsection{two examples of toric singularities}

Here, we give two examples of toric manifolds and toric singularities as a
way to illustrate the general idea. We first consider the simple case of $%
\mathbb{C}$, the set of complex numbers; then we describe the case of the
complex projective line $CP^{1}$. The latter is one of the basic objects in
dealing with toric geometry; in particular in building toric graphs and
studying toric singularities.

$\mathbf{1)}$ \emph{the complex line} $\mathbb{C}\sim \mathbb{R}^{2}$\newline
The complex line $\mathbb{C}$ is the simplest example of toric manifolds. To
exhibit the toric structure of $\mathbb{C}$ and build the corresponding
toric graph \textrm{\cite{D0,E1,E2,E3}}, it is convenient to think about the
complex coordinate $z=x+iy$ as 
\begin{equation}
z=\left \vert z\right \vert e^{i\varphi}
\end{equation}
with $\left \vert z\right \vert =\sqrt{x^{2}+y^{2}}\in$ $\mathbb{R}^{+}$ and 
$\tan \varphi=\frac{y}{x}$. By using this representation, the complex line $%
\mathbb{C}$ may be viewed as the fibration of a circle $\mathbb{S}^{1}$ over
the real half line $\mathbb{R}^{+}$; i.e: 
\begin{equation}
\mathbb{C\sim R}^{+}\times \mathbb{S}^{1}
\end{equation}
This non compact complex line admits a natural $U\left( 1\right) $ action
operating as%
\begin{equation}
z\rightarrow z^{\prime}=ze^{i\theta}\qquad \Leftrightarrow \qquad \varphi
\rightarrow \varphi+\theta
\end{equation}
with a fixed point at $z=0$. In toric geometry, the complex line $\mathbb{C}$
is represented by a toric graph given by the half-line $\mathbb{R}^{+}=\left[
0,\infty \right[ $, corresponding to $\left \vert z\right \vert $, above
which there is a circle of radius $r=r\left( \left \vert z\right \vert
\right) $. 
\begin{figure}[ptbh]
\begin{center}
\hspace{0cm} \includegraphics[width=8cm]{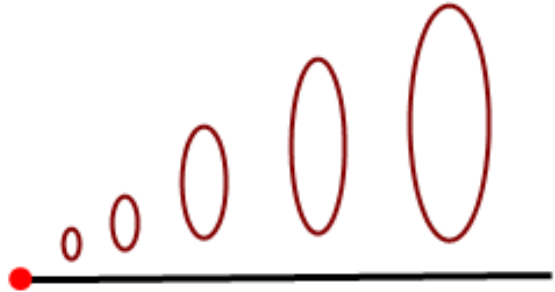}
\end{center}
\par
\vspace{-1cm}
\caption{toric graph of $\mathbb{C}$ viewed as a half-line $\mathbb{R}^{+}=%
\left[ 0,\infty \right[ $ with a circle on top which shrinks at the origin.}
\label{TA}
\end{figure}
The toric graph of $\mathbb{C}$ is depicted in fig \ref{TA}; the circle of
the fibration shrinks to zero at the end of the half-line where live the fix
point of the toric action; 
\begin{equation}
z=z^{\prime}=0.  \label{x}
\end{equation}

\begin{figure}[ptbh]
\begin{center}
\hspace{0cm} \includegraphics[width=8cm]{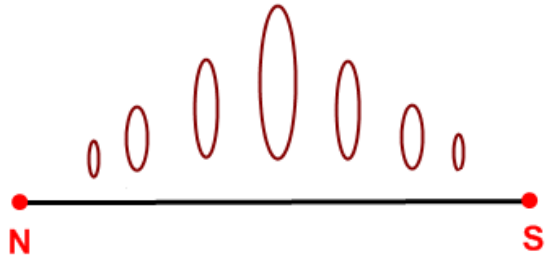}
\end{center}
\par
\vspace{-1cm}
\caption{toric graph of $CP^{1}\sim$ $\mathbb{S}^{2}$ which can be viewed as
an interval $\left[ N,S\right] $ with a circle on top. The circle fiber
shrinks to zero size at the ends N and S.}
\label{TB}
\end{figure}

$\mathbf{2)}$ \emph{the projective line }$CP^{1}\sim \mathbb{S}^{2}$\newline
The projective line $CP^{1}$ is one of the basic objects in drawing toric
graphs of toric manifolds; it is then useful to have some extensive details
on this compact complex curve, which in the real geometry language,
describes precisely the real 2-sphere $\mathbb{S}^{2}$ considered in section
3.1.2. First notice that $\mathbb{S}^{2}$ may be realized in various, but
equivalent, manners depending on the targeted features. We have the three
following realizations of $\mathbb{S}^{2}$: \newline
$\left( i\right) $ by embedding it in the real 3-euclidian space as usual
like $\left( x_{1}\right) ^{2}+\left( x_{2}\right) ^{2}+\left( x_{3}\right)
^{2}=\varrho^{2}$ with $x_{i}$ standing for the coordinates of $\mathbb{R}%
^{3}$. \newline
$\left( ii\right) $ by thinking about it as the coset group manifold 
\begin{equation}
\mathbb{S}^{2}\sim SU\left( 2\right) /U\left( 1\right)
\end{equation}
which is realized by using two complex variables $w_{+},w_{-}$ constrained
as \textrm{\cite{J1,J2}} 
\begin{equation}
\left \vert w_{+}\right \vert ^{2}+\left \vert w_{-}\right \vert ^{2}=\xi
,\qquad \xi \geq0,  \label{w}
\end{equation}
together with the gauge identification%
\begin{equation}
w_{+}\equiv e^{i\varphi}w_{+},\qquad w_{-}\equiv e^{-i\varphi}w_{-}
\end{equation}
Equation (\ref{w}) fixes one degree of the real 4 degrees of freedom leaving
three free ones; the second relation reduces this number down to 2. The real
parameter $\xi$ in (\ref{w}) is the Kahler parameter of the real 2-sphere;
it controlls its volume. Notice also that the variables $\left(
w_{+},w_{-}\right) $ form two states of an $SU\left( 2\right) $ doublet with
opposite $U\left( 1\right) $ charges as shown on the following relation%
\begin{equation}
\left( 
\begin{array}{c}
w_{+} \\ 
w_{-}%
\end{array}
\right) \rightarrow e^{i\varphi \tau^{3}}\left( 
\begin{array}{c}
w_{+} \\ 
w_{-}%
\end{array}
\right) ,\qquad \tau^{3}=\left( 
\begin{array}{cc}
1 & 0 \\ 
0 & -1%
\end{array}
\right)
\end{equation}
This feature can be seen by setting 
\begin{equation}
\begin{tabular}{lll}
$w^{\alpha}=\left( 
\begin{array}{c}
w_{+} \\ 
w_{-}%
\end{array}
\right) $ & $,$ & $w_{\alpha}=\left( 
\begin{array}{c}
w_{-} \\ 
-w_{+}%
\end{array}
\right) $ \\ 
&  &  \\ 
$\bar{w}_{\alpha}=\left( 
\begin{array}{c}
\bar{w}_{+} \\ 
\bar{w}_{-}%
\end{array}
\right) $ & $,$ & $\bar{w}^{\alpha}=\left( 
\begin{array}{c}
-\bar{w}_{-} \\ 
\bar{w}_{+}%
\end{array}
\right) $%
\end{tabular}%
\end{equation}
with $\bar{w}_{\alpha}=\overline{\left( w^{\alpha}\right) }$ where the index 
$\alpha$ is raised and lowered by help of the antisymmetric $2\times2 $
metric tensor $\varepsilon_{\alpha \beta}$ of spinors. As such, we have 
\begin{equation}
w^{\alpha}\bar{w}_{\alpha}=\varepsilon_{\alpha \beta}w^{\alpha}\bar{w}%
^{\beta}
\end{equation}
which upon using the above relations leads to precisely $\left \vert
w_{+}\right \vert ^{2}+\left \vert w_{-}\right \vert ^{2}$ showing that it
is invariant under $SU\left( 2\right) $ transformations. We also have $%
w^{\alpha}w_{\alpha}=\varepsilon_{\alpha \beta}w^{\alpha}w^{\beta}=0$ and
similarly for the complex conjugates. \newline
$\left( iii\right) $ the 2-sphere $\mathbb{S}^{2}$ may be as well realized
in terms of the projective line $CP^{1}$ which is given by the
compactification of the complex line. We have \textrm{\cite{J2}} 
\begin{equation}
CP^{1}=\mathbb{C}^{2}/\mathbb{C}^{\ast}
\end{equation}
with $\mathbb{C}^{\ast}$ standing for the symmetry action $\left( z,\zeta
\right) \rightarrow \left( \lambda z,\lambda \zeta \right) $. The toric
graph of the projective line is given by fig \ref{TB}. 
The projective line has also a $U\left( 1\right) $ toric action in terms of
which the 2-sphere can be represented as an interval $\left[ N,S\right] $
times a circle $\mathbb{S}^{1}$ that shrinks at the two ends N and S,
corresponding to north and south poles of the sphere. \newline
The coordinate $\varsigma$ on the interval is a function of $\left \vert
z\right \vert $, which upon using the Fubini-Study metric, reads as \textrm{%
\cite{E1}}, 
\begin{equation}
\varsigma=\frac{\eta \left \vert z\right \vert ^{2}}{1+\left \vert z\right
\vert ^{2}}  \label{y}
\end{equation}
and runs form $0$ to $\eta$ defining the length of the interval $\left[ N,S%
\right] $ and so the size of the 2-sphere. Notice that in dealing with $%
CP^{1}$; one has to distinguish between the homogeneous coordinates\footnote{%
In eq(\ref{cp}) the homogenous $\left( Z_{1},Z_{0}\right) $ are denoted as $%
\left( z_{1},\zeta_{1}\right) $} $\left( Z_{1},Z_{0}\right) $ and the affine
ones that corresponds to working in a local patch.

\subsection{the complex surface\emph{\ }$CP^{1}\times CP^{1}$}

\begin{figure}[ptbh]
\begin{center}
\hspace{0cm} \includegraphics[width=8cm]{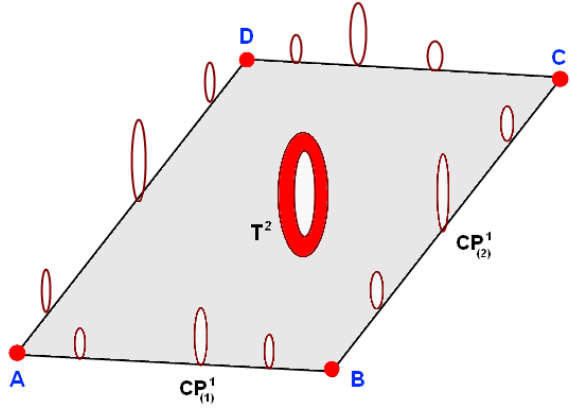}
\end{center}
\par
\vspace{0cm}
\caption{toric graph of the complex surface $CP^{1}\times CP^{1}\sim$ $%
\mathbb{S}^{2}\times \mathbb{S}^{2}$. On the edges a 1-cycle of the 2-torus
shrinks leaving $\mathbb{S}^{1}.$ At the 4 vertices both the 1-cycles of T$%
^{2}$ shrink to zero; these points correspond to fix points of the $%
U^{2}\left( 1\right) $ toric action.}
\label{OK}
\end{figure}

This is a complex 2- (real 4-) dimensional compact manifold that can be
described by 4 complex homogeneous variables%
\begin{equation}
\left( z_{1},\zeta_{1},z_{2},\zeta_{2}\right) \text{ \ }\in \text{ \ }\frac{%
\mathbb{C}^{2}\times \mathbb{C}^{2}}{\mathbb{C}^{\ast}\times \mathbb{C}%
^{\ast}}
\end{equation}
obeying the identifications 
\begin{equation}
\begin{tabular}{lll}
$\left( z_{1},\zeta_{1}\right) $ & $\qquad \rightarrow \qquad$ & $\left(
\lambda z_{1},\lambda \zeta_{1}\right) $ \\ 
$\left( z_{2},\zeta_{2}\right) $ & $\qquad \rightarrow \qquad$ & $\left( \mu
z_{2},\mu \zeta_{2}\right) $%
\end{tabular}%
\end{equation}
where $\lambda$ and $\mu$ are two non zero complex numbers. This complex
surface has a $U^{2}\left( 1\right) $ toric action, consisting of the $%
U^{4}\left( 1\right) $ action on the phases of $\left( z_{i},\zeta
_{i}\right) $ modulo the action of the diagonal $U\left( 1\right) \times
U\left( 1\right) $ which act trivially on $CP^{1}\times CP^{1}$. Being a
toric manifold, $CP^{1}\times CP^{1}$ may be viewed as the fibration%
\begin{equation}
CP^{1}\times CP^{1}\sim \Delta_{2}\times T^{2}
\end{equation}
with real 2-dimensional base $\Delta_{2}$ and a fiber $T^{2}$. The toric
diagram of this manifold is given by the tensor product of the toric graphs
of the two $CP^{1}$'s; the resulting polytope is as in fig \ref{OK}; the
base $\Delta_{2}$ is a parallelogram $\left[ ABCD\right] $ with vertices 
\begin{equation}
\begin{tabular}{lll}
$A=\left( 0,1,0,1\right) $ & $,$ & $C=\left( 1,0,0,0\right) $ \\ 
$B=\left( 1,0,0,1\right) $ & $,$ & $D=\left( 1,0,1,0\right) $%
\end{tabular}%
\end{equation}
and should be associated with the unit cell in the reciprocal space given by
fig \ref{CU}. Let us describe rapidly the construction of this toric graph.
In the coordinate patch where $\zeta_{1}=\zeta_{2}=1$, we can consider a
basis of the $U^{2}\left( 1\right) $ action to consist of 
\begin{equation}
\left( z_{1},1,z_{2},1\right) \qquad \rightarrow \qquad \left( z_{1}^{\prime
},1,z_{2}^{\prime},1\right) =\left(
z_{1}e^{i\theta_{1}},1,z_{2}e^{i\theta_{2}},1\right)
\end{equation}
The fixed point of the $\theta_{1}$ action consists of the complex line $%
\left( 0,1,z_{2},1\right) $ describing a local coordinate patch of a
projective line $CP^{1}$ to which we refer as the copy $CP_{\left( 2\right)
}^{1}$ parameterized by $\frac{z_{2}}{\zeta_{2}}=z_{2}$. Similarly the fixed
point of the $\theta_{2}$ action is given by $\left( z_{1},1,0,1\right) $;
it is a local patch of a $CP_{\left( 1\right) }^{1}$ parameterized by $\frac{%
z_{1}}{\zeta_{1}}=z_{1}$. The intersection of the two lines is given by the
point 
\begin{equation}
\left( 0,1,0,1\right)
\end{equation}
which is a fix point of both $\theta_{1}$ and $\theta_{2}$\ actions. The
same situation takes place for the other patches; for instance if we work in
the coordinate patch where $z_{1}=\zeta_{2}=1$, we can consider a basis of
the $U^{2}\left( 1\right) $ action to consist of%
\begin{equation}
\left( 1,\zeta_{1},z_{2},1\right) \qquad \rightarrow \qquad \left( 1,\zeta
_{1}^{\prime},z_{2}^{\prime},1\right) =\left(
1,\zeta_{1}e^{i\phi_{1}},z_{2}e^{i\theta_{2}},1\right)
\end{equation}
The fix point of the $\phi_{1}$ action consists of $\left(
1,0,z_{2},1\right) $ and describes the projective line $CP_{\left( 2\right)
}^{1}$ parameterized by $\frac{z_{2}}{\zeta_{2}}=z_{2}$. While the fixed
point of the $\theta_{2}$ action is a $CP^{1}$ parameterized by $\frac{z_{1}%
}{\zeta_{1}}=\frac{1}{\zeta_{1}}$; it is precisely the $CP_{\left( 1\right)
}^{1}$; but taken in the local coordinate patch $\left( 1,\zeta_{1}\right) $.

\subsection{Creutz Flavors}

First recall that for QCD$_{2}$ naive fermions, the Dirac operator $D_{NF}=$ 
$i\gamma_{1}\sin p_{1}+$ $i\gamma_{2}\sin p_{2}$ has four zeros in
reciprocal space located at $\mathbf{P}_{m_{1},m_{2}}=(P_{1}^{\left(
m_{1}\right) },P_{2}^{\left( m_{2}\right) })$ with:%
\begin{equation}
\begin{tabular}{lll}
$P_{1}^{\left( m_{1}\right) }=m_{1}\pi,$ & $P_{2}^{\left( m_{2}\right)
}=m_{2}\pi,$ & $m_{l}=0,1$%
\end{tabular}
\label{4Z}
\end{equation}
These 4 zeros are also the fix points of the $\mathbb{Z}_{2}$ anti-symmetry
of $D_{NF}$ 
\begin{equation}
\left( p_{1},p_{2}\right) \rightarrow \left( -p_{1},-p_{2}\right)
\end{equation}
they form the vertices of a unit cell in the reciprocal space as depicted in
fig \ref{CU}. At these 4 vertices of the unit cell which we denote as%
\begin{equation}
\begin{tabular}{lll}
$A=(P_{1}^{\left( 0\right) },P_{2}^{\left( 0\right) })$ & $,$ & $%
B=(P_{1}^{\left( 1\right) },P_{2}^{\left( 0\right) })$ \\ 
$C=(P_{1}^{\left( 0\right) },P_{2}^{\left( 1\right) })$ & $,$ & $%
D=(P_{1}^{\left( 1\right) },P_{2}^{\left( 1\right) })$%
\end{tabular}
\label{v}
\end{equation}
both the Dirac operator coefficients $\sin p_{1}$ and $\sin p_{2}$ vanish;
while on the 4 edges%
\begin{equation}
\begin{tabular}{llll}
$\left[ AB\right] ,$ & $\left[ BC\right] ,$ & $\left[ CD\right] ,$ & $\left[
DA\right] $%
\end{tabular}%
\end{equation}
only one of the two $\sin p_{l}$'s of $D_{naive}$ that vanishes. Moreover,
inside the surface of the cell, $\sin p_{1}$ and $\sin p_{2}$ are both of
them non zero exactly as in for the toric graph of 
\begin{equation*}
CP^{1}\times CP^{1}.
\end{equation*}
Now, expanding $D_{naive}$ near each one of the 4 zeros (\ref{v}) by setting 
$p_{l}=P_{l}^{\left( m_{l}\right) }+q_{l}$ and using the identity 
\begin{equation}
\sin \left( q_{l}+P_{l}^{\left( m_{l}\right) }\right) =\sin q_{l}\cos
P_{l}^{\left( m_{l}\right) }+\cos q_{l}\sin P_{l}^{\left( m_{l}\right) }
\end{equation}
we get 4 kinds of Dirac operators $D_{\left( m_{1},m_{2}\right) }$ in the
continuum. These expansions read collectively as%
\begin{equation}
D_{\left( m_{1},m_{2}\right) }\simeq \gamma_{\left( m_{1}\right)
}^{1}q_{1}+\gamma_{\left( m_{2}\right) }^{2}q_{2}+O\left( q^{2}\right)
\end{equation}
with $\gamma_{\left( m_{1}\right) }^{1}$ and $\gamma_{\left( m_{2}\right)
}^{2}$ related to the standard ones $\gamma^{l}$ like, 
\begin{equation}
\gamma_{\left( m_{1}\right) }^{1}=\left( -\right) ^{m_{1}}\gamma ^{1},\qquad
\gamma_{\left( m_{2}\right) }^{2}=\left( -\right) ^{m_{2}}\gamma^{2}
\end{equation}
and obeying as well the same Clifford algebra. The above relations can be
also written in terms of the following similarity transformations%
\begin{equation}
\gamma_{\left( m_{1}\right) }^{1}=\Gamma_{\left( m_{1},0\right)
}^{+}\gamma^{1}\Gamma_{\left( m_{1},0\right) },\qquad \gamma_{\left(
m_{2}\right) }^{2}=\Gamma_{\left( 0,m_{2}\right)
}^{+}\gamma^{2}\Gamma_{\left( 0,m_{2}\right) }
\end{equation}
with%
\begin{equation}
\Gamma_{\left( m_{1},m_{2}\right) }=\left( \gamma^{1}\right) ^{m_{2}}\left(
\gamma^{2}\right) ^{m_{1}}
\end{equation}
Applying the point splitting method to this model by identifying the 4 \emph{%
inequivalent} species associated with the operators $D_{\left(
m_{1},m_{2}\right) }$ as 4 independent flavors, we end with the 4 Creutz
flavors%
\begin{equation}
\psi_{\left( m_{1}m_{2}\right) }=\left( 
\begin{array}{cc}
\psi_{\left( 00\right) } & \psi_{\left( 01\right) } \\ 
\psi_{\left( 10\right) } & \psi_{\left( 11\right) }%
\end{array}
\right)
\end{equation}
with%
\begin{equation}
\psi_{\left( m_{1}m_{2}\right) }\text{ \ }\sim \text{ \ }\Gamma_{\left(
m_{1}m_{2}\right) }\psi \left( p_{1}-P_{1}^{\left( m_{1}\right)
},p_{2}-P_{2}^{\left( m_{2}\right) }\right)
\end{equation}
Moreover, thinking about $\sin p_{1}$ and $\sin p_{2}$ as the radii of two
circles as in eq(\ref{313}); it follows that the unit cell of fig \ref{CU}
is precisely the real base of the toric graph of $CP^{1}\times CP^{1}$ given
by fig \ref{OK}.\newline
Furthermore, rewriting the 4 zeros (\ref{4Z})%
\begin{equation}
\begin{tabular}{lllll}
$\left( P_{1},P_{2}\right) =$ & $\left( 0,0\right) ,$ & $\left( \pi,0\right) 
$, & $\left( \pi,0\right) ,$ & $\left( \pi,\pi \right) $%
\end{tabular}%
\end{equation}
in terms of the wave vectors $\left( K_{1},K_{2}\right) =\left(
bP_{1},bP_{2}\right) $, with $b$ given by the inverse of the real lattice
spacing parameter ($b=\frac{1}{a}$), we obtain the following parallelogram 
\begin{equation}
\begin{tabular}{lllll}
$\left( K_{1},K_{2}\right) =$ & $\left( 0,0\right) ,$ & $\left(
b\pi,0\right) $, & $\left( b\pi,0\right) ,$ & $\left( b\pi,b\pi \right) $%
\end{tabular}%
\end{equation}
with area%
\begin{equation}
\mathcal{A}=\pi^{2}b^{2}
\end{equation}
Notice that in the infrared limit $b\rightarrow0$, the area of the unit cell
in reciprocal lattice shrinks and the 4 zeros collide leading to an $%
SU\left( 2\right) \times SU^{\prime}\left( 2\right) $ singularity. This
means that the Creutz multiplet $\psi_{\left( m_{1}m_{2}\right) }$
transforms in the $\left( \frac{1}{2},\frac{1}{2}\right) $ representation of 
$SU\left( 2\right) \times SU^{\prime}\left( 2\right) $.

\section{KW fermions and Conifold}

We begin by recalling the Dirac operator $D_{KW}$ of the KW fermions on
4-dimensional lattice. This is anti-hermitian 4$\times$4 matrix operator
given by (\ref{ope}) which, for convenience, we rewrite it as 
\begin{equation}
D_{{\small KW}}=\dsum \limits_{l=1}^{4}i\mathbf{\gamma}_{l}F_{l}
\end{equation}
with $F_{l}=\sin p_{l}$ for $l=1,2,3$ as for naive fermions; and the fourth
component $F_{4}$ given by 
\begin{equation}
F_{4}=\frac{1}{\sin \alpha}\left( \cos \alpha+3-\sum_{l=1}^{4}\cos
p_{l}\right)  \label{F}
\end{equation}
Generally, this operator depends on three basics objects namely: $\left(
1\right) $ the \emph{4} gamma matrices $\gamma_{1},$ $\gamma_{2},$ $%
\gamma_{3},$ $\gamma_{4}$ satisfying the euclidian Clifford algebra $%
\gamma_{\mu}\gamma_{\nu}+\gamma_{\nu}\gamma_{\mu}=2\delta_{\mu \nu}$ leading
to the remarkable property 
\begin{equation}
\left( D_{{\small KW}}\right) ^{2}=-I_{4}\left( \dsum
\limits_{l=1}^{4}F_{l}^{2}\right) ,\qquad D_{{\small KW}}D_{{\small KW}%
}^{+}=-\left( D_{{\small KW}}\right) ^{2}
\end{equation}
where $I_{4}$ is the 4$\times$4 identity matrix. $\left( 2\right) $ the real
4 phases $e^{ip_{1}},$ $e^{ip_{2}},$ $e^{ip_{3}},$ $e^{ip_{4}}$ of the wave
functions of the particle propagating along the directions towards the first
nearest neighbors in the real 4d-hypercubic lattice; and $\left( 3\right) $
the free real parameter $\alpha$ showing that the KW operator (\ref{ope}) is
in fact a one- parameter flow operator%
\begin{equation}
D_{{\small KW}}=D\left( \alpha \right) .
\end{equation}
with spectral parameter $\alpha$.

\subsection{zeros of $D_{KW}$ as fix points of $\mathbb{Z}_{2}$ symmetry}

The operator $D_{KW}$ has some remarkable features that allows to interpret
the point splitting method of Creutz in terms of a blown up singularity in
complex geometry. Four useful features for this study are collected below: 
\newline
$i)$ $D_{KW}$ is a periodic operator in the reciprocal space variables $%
p_{1},p_{2},p_{3},p_{4}$ and also in the spectral parameter $\alpha$. So the
variation of the $p_{l}$'s and $\alpha$ can be restricted to a particular
period which can be taken as 
\begin{equation}
\begin{tabular}{llll}
$p_{l},\alpha \in \left] -\pi,\pi \right] $ & , & $\func{mod}2\pi$ & 
\end{tabular}
\label{ph}
\end{equation}
$ii)$ $D_{KW}$ is moreover odd under the $\mathbb{Z}_{2}$ parity
transformation reversing simultaneously the sign of the phase variables $%
p_{l}$ and the spectral parameter $\alpha$ as follows%
\begin{equation}
\left( p_{1},p_{2},p_{3},p_{4};\alpha \right) \rightarrow \left(
-p_{1},-p_{2},-p_{3},-p_{4};-\alpha \right)  \label{z}
\end{equation}
This $\mathbb{Z}_{2}$ anti-symmetry allows to restrict further the domain (%
\ref{ph}) down to%
\begin{equation}
p_{l},\alpha \in \left[ 0,\pi \right]
\end{equation}
We will refer to this fundamental domain in the reciprocal space as the unit
cell; this is a 4-dimensional hypercube with volume, in terms of the spacing
parameter $a$ of the real lattice, given by $\frac{\pi^{4}}{a^{4}}$.\newline
$iii)$ the operator $D_{{\small KW}}$ admits also another property that has
no analogue in the case of naive fermions; it is singular for $%
\alpha=\alpha_{sg}$ with 
\begin{equation}
\alpha_{sg}=0,\text{ }\pi,\qquad \func{mod}2\pi
\end{equation}
as explicitly shown on the expression of $F_{4}$ (\ref{F}). This feature can
be also exhibited by computing the leading terms of the expansion of $F_{4}$
for $\alpha$ close to $\alpha_{sg}$. In the case $\alpha \sim0$, we have 
\begin{equation}
F_{4}\simeq \frac{1}{\alpha}\left( 4-\sum_{l=1}^{4}\cos p_{l}\right) +\frac{%
\alpha}{6}\left( 1-\sum_{l=1}^{4}\cos p_{l}\right) +O\left( \alpha^{3}\right)
\end{equation}
whose leading term has a pole at $\alpha=0$. Notice that at this pole, the
function $F_{4}$ diverges as far as $\sum_{l=1}^{4}\left( 1-\cos
p_{l}\right) \neq0$.\  \newline
$iv)$ In the case where $\alpha \neq \alpha_{sg}$, the operator $D_{{\small %
KW}}$ has two simple zero modes located at 
\begin{equation}
\mathbf{P}_{\pm}=\left( 0,0,0,\pm \alpha \right) =\pm \alpha \mathbf{e}_{4}
\end{equation}
with $\mathbf{e}_{4}=\left( 0,0,0,1\right) $ standing for the fourth
direction of the reciprocal space. These zeros, which are related by the $%
\mathbb{Z}_{2}$ anti-symmetry (\ref{z}), show that:

\begin{itemize}
\item Karsten-Wilczek fermions have non trivial wave phases only along the $%
\mathbf{e}_{4}$- direction,

\item the two zeros collide in the limit $\alpha \rightarrow0$ giving a rank
two degenerate zero: 
\begin{equation}
\lim_{\alpha \rightarrow0}\mathbf{P}_{+}=\lim_{\alpha \rightarrow0}\mathbf{P}%
_{-}=\left( 0,0,0,0\right)
\end{equation}
\end{itemize}

\  \  \newline
Observe also that for those $p_{l}$ phases that are in the nearby of the
zeros $\mathbf{P}_{\pm}$, say for instance $p_{l}=\mathbf{P}_{+}+q_{l}$ with
small $q_{l}$, the coefficient $F_{4}$ reads as 
\begin{equation}
F_{4}=\frac{1}{\sin \alpha}\left( 3-\sum_{l=1}^{3}\cos q_{l}+\cos \alpha
\left( 1-\cos q_{4}\right) +\sin \alpha \sin q_{4}\right)
\end{equation}
and expands, at first order in $q$, like $F_{4}\simeq q_{4}+O\left(
q_{l}^{2},\alpha \right) $.

\subsection{$F_{4}$ as a resolved conifold singularity}

By writing the Dirac operator as%
\begin{equation}
D_{KW}=+i\mathbf{\gamma}_{1}\sin p_{1}+i\mathbf{\gamma}_{2}\sin p_{2}+i%
\mathbf{\gamma}_{3}\sin p_{3}+\frac{i}{\sin \alpha}\mathcal{F}_{4}
\end{equation}
with $\mathcal{F}_{4}$ given by the $\alpha$- dependent function 
\begin{equation}
\mathcal{F}_{4}\left( \alpha \right) =\left( 4-\sum_{l=1}^{4}\cos
p_{l}\right) -\left( 1-\cos \alpha \right)  \label{F4}
\end{equation}
we learn that the zeros of the $\sin p_{l}$ coefficients along the $\mathbf{%
\gamma}_{1},$ $\mathbf{\gamma}_{2},$ $\mathbf{\gamma}_{3}$ directions are as
in the case of naive fermions and so are interpreted in terms of toric
singularities as in fig \ref{OK}. However the two simple zeros of the term $%
\mathcal{F}_{4}$ have a different geometric interpretation; they are
associated with the small resolution of the conifold singularity which we
prove hereafter.

\subsubsection{useful tools on singularities}

To get the link between the two zeros of the KW fermions and the small
resolution of conifold singularity, it is helpful to start by giving some
useful tools on $SU\left( 2\right) $ and conifold singularities; then turn
back to derive the relation with zeros of KW fermions. To this end we will
proceed as follows:

\begin{itemize}
\item first, we describe briefly the $SU\left( 2\right) $ singularity and
its resolution in terms of a blown up real 2-sphere where the order 2
degenerate zero gets replaced by two simple zeros \textrm{\cite{C1,C2}}.

\item second, we study the conifold singularity together with its complex
and Kahler deformations. The latter, known also as the small resolution,
uses a blown up 2-sphere to lift the conifold singularity \textrm{\cite%
{G1,G2,G3,G4}}; it is the one related to the zeros of KW fermions.
\end{itemize}

\begin{figure}[ptbh]
\begin{center}
\hspace{0cm} \includegraphics[width=14cm]{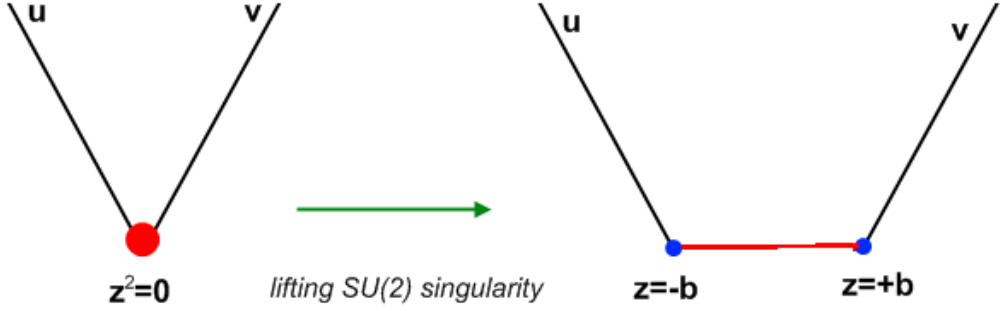}
\end{center}
\par
\vspace{0cm}
\caption{on left the SU$\left( 2\right) $ singularity at the origin. On
right its resolution; the singular point $z=0$ has been replaced by a
complex projective line.}
\label{A1}
\end{figure}

\textbf{A)} $SU\left( 2\right) $\emph{\ singularity }\newline
There are various ways to introduce this singularity; the most natural one
is given by considering those singular complex surfaces 
\begin{equation}
\mathcal{G}\left( z_{1},z_{2},z_{3}\right) =0
\end{equation}
embedded in $\mathbb{C}^{3}$ and known as the Asymptotic Locally Euclidian
(ALE) space. In this case, the shape of the complex surface $\mathcal{G}$
near the singularity is described by the following complex algebraic
relation 
\begin{equation}
\mathcal{G}\left( z_{1},z_{2},z_{3}\right) =\left( z_{1}\right) ^{2}+\left(
z_{2}\right) ^{2}+\left( z_{3}\right) ^{2}  \label{sin}
\end{equation}
which is invariant under the $\mathbb{Z}_{2}$ symmetry $z_{l}\rightarrow
z_{l}^{\prime}=-z_{l}$. This discrete symmetry has a fix point $%
z_{l}^{\prime }=z_{l}$ at the origin $\left( z_{1},z_{2},z_{3}\right)
=\left( 0,0,0\right) $ where live indeed an $SU\left( 2\right) $
singularity. Up on performing the variable change 
\begin{equation}
u=z_{1}+iz_{2},\qquad v=z_{1}-iz_{2},\qquad z=z_{3}
\end{equation}
we can bring the above relation to the familiar form, 
\begin{equation}
uv=z^{2}
\end{equation}
with fix point of the $\mathbb{Z}_{2}$ symmetry at $\left( u,v,z\right)
=\left( 0,0,0\right) $. Recall that by $SU\left( 2\right) $ singularity, we
mean that the equation $\mathcal{G}=0$ and its differential $d\mathcal{G}=0$
have the same zeros which in this example is precisely the fix point of the $%
\mathbb{Z}_{2}$ symmetry namely $\left( z_{1},z_{2},z_{3}\right) =\left(
0,0,0\right) $. \newline
Notice also that the complex surface (\ref{sin}) is a manifold belonging to
an interesting class of Kahler manifolds known as \emph{toric manifolds}.
These manifolds, introduced in section 4, admit toric actions and are nicely
represented by toric diagrams. In the case of the ALE space with SU$\left(
2\right) $ singularity, the corresponding toric graph and its deformation
are as in fig \ref{A1}.

From the structure of eq(\ref{sin}), the complex 2- dimension surface $%
\mathcal{G}\left( z_{1},z_{2},z_{3}\right) =0$ is nothing but the tangent
bundle of the 2-sphere $T^{\ast}\mathbb{S}^{2}$ which may be thought of as
given by the fibration $\mathbb{S}^{2}\times \mathbb{R}^{2}$. This feature
can be viewed by working with the real coordinates; by using $%
z_{l}=x_{l}+iy_{l}$ and putting back into (\ref{sin}), we obtain the two
real equations 
\begin{equation}
\begin{tabular}{lll}
$\dsum \limits_{l=1}^{3}\left( x_{l}^{2}-y_{l}^{2}\right) =0$ & , & $\dsum
\limits_{l=1}^{3}x_{l}y_{l}=0$%
\end{tabular}%
\end{equation}
describing the fibration $T^{\ast}\mathbb{S}^{2}\simeq \mathbb{S}^{2}\times 
\mathbb{R}^{2}$.\ The compact part of this manifold obtained by setting $%
y_{l}=0$; the reduced relation describes a real 2-sphere $\mathbb{S}^{2}$;
but with vanishing volume; that is a singular 2-sphere:%
\begin{equation}
\dsum \limits_{l=1}^{3}x_{l}^{2}=0
\end{equation}
The singular surface (\ref{sin}) may be smoothed by giving a non zero volume
to the above singular $\mathbb{S}^{2}$; this is achieved by replacing (\ref%
{sin}) by the deformed relation 
\begin{equation}
z_{1}^{2}+z_{2}^{2}+z_{3}^{2}=\beta^{2}
\end{equation}
where $\beta$ is an arbitrary non zero number. Rewriting this deformed
relation as%
\begin{equation}
z_{1}^{2}+z_{2}^{2}+\left( z_{3}-\beta \right) \left( z_{3}+\beta \right) =0
\end{equation}
one sees that the double zero at $z_{3}^{2}=0$ is now splited into two
simple zeros located at%
\begin{equation}
z_{3}=+\beta,\qquad z_{3}=-\beta
\end{equation}
This is almost what happens with the zeros of the KW fermions which are
located in the real 4-dimensional reciprocal space at 
\begin{equation}
p_{4}=+\alpha,\qquad p_{4}=-\alpha
\end{equation}
the role of the parameter $\beta$ is played by the free parameter $\alpha$.
We will turn to this relation in next subsection; in due time notice that in
toric geometry language, the SU$\left( 2\right) $ singularity of the ALE
space corresponds to the shrinking of 2-torus at the origin; the deformation
corresponds to blowing up a 2-sphere at the singularity. So the two
degenerate zeros living at the fix point get splited and pushed towards the
north and south poles of the blown up sphere as in fig \ref{TB}. The blown
up breaks the $SU\left( 2\right) $ living at the singularity down to its $%
U\left( 1\right) $ subgroup%
\begin{equation}
SU\left( 2\right) \rightarrow U\left( 1\right)
\end{equation}

\begin{figure}[ptbh]
\begin{center}
\hspace{0cm} \includegraphics[width=14cm]{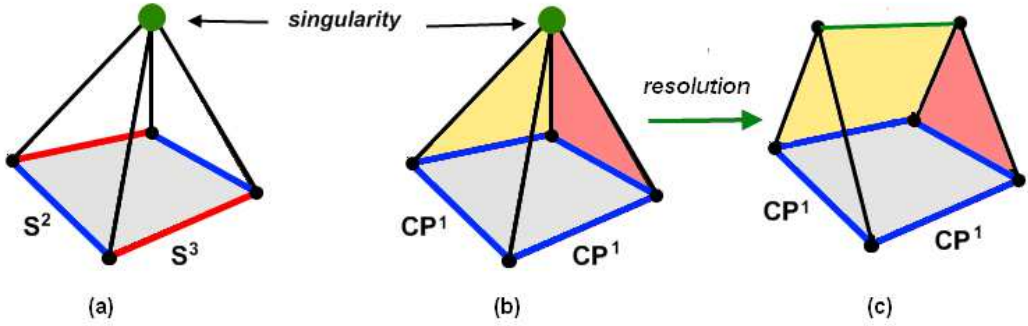}
\end{center}
\par
\vspace{0cm}
\caption{(a) the conifold viewed as $\mathbb{S}^{2}\times \mathbb{S}%
^{3}\times \mathbb{R}^{+}$; the resolution of the singularity can be
achieved in 2 ways, either by blowing a 3-sphere or a 2-sphere. (b) the
conifold viewed as the toric diagram of $CP^{1}\times CP^{1}\times \mathbb{C}
$. (c) Small resolution of conifold by blowing up the singularity by a $%
CP^{1}$ curve.}
\label{OC}
\end{figure}

\textbf{B)} \emph{3d- conifold singularity} \newline
The complex 3d-conifold we are interested in here is given by the complex
3-dimensional affine variety generally defined by the complex algebraic
equation 
\begin{equation}
G\left( u,v,z,w\right) =uv-zw=0  \label{uv}
\end{equation}
describing a singular complex 3-dimensional hypersurface embedded in $%
\mathbb{C}^{4}$. By using the following change of variables 
\begin{equation}
\begin{tabular}{lllll}
$u=z_{1}+iz_{2}$ & $,$ & $v=z_{1}-iz_{2}$ & , & $w=-z_{4}+iz_{3}$ \\ 
$z=z_{1}+iz_{2}$ & , & $z=z_{4}+iz_{3}$ &  & 
\end{tabular}%
\end{equation}
this complex relation can be brought to the diagonal form%
\begin{equation}
\left( z_{1}\right) ^{2}+\left( z_{2}\right) ^{2}+\left( z_{3}\right)
^{2}+\left( z_{4}\right) ^{2}=0.  \label{A2}
\end{equation}
Viewed as a real 6-dimensional affine variety embedded into the euclidian
space $\mathbb{R}^{8}$, the real relations that define the conifold are
obtained by substituting $z_{l}=x_{l}+iy_{l}$ back into (\ref{A2}); this
gives 
\begin{equation}
\begin{tabular}{lll}
$\dsum \limits_{l=1}^{4}\left( x_{l}^{2}-y_{l}^{2}\right) =0$ & , & $\dsum
\limits_{l=1}^{4}x_{l}y_{l}=0$%
\end{tabular}
\label{2A}
\end{equation}
with compact part given by the singular real 3-sphere $\mathbb{S}^{3}$%
\begin{equation}
\dsum \limits_{l=1}^{4}x_{l}^{2}=0
\end{equation}
The non compact part, parameterized by the $y_{l}$ variables, is given by
the real 3-dimensional space $\mathbb{R}^{3}$ which, in spherical
coordinates, can be also viewed as given by the fibration of a real 2-sphere
on the half line like $\mathbb{R}^{+}\times \mathbb{S}^{2}$. So the conifold
may be thought of as 
\begin{equation}
\mathbb{R}^{+}\times \mathbb{S}^{2}\times \mathbb{S}^{3}
\end{equation}
The geometry of this real 6-dimensional conifold can be therefore imagined
as a cone in $\mathbb{R}^{8}$ with base $\mathbb{S}^{2}\times \mathbb{S}^{3}$
and top at the origin as heuristically depicted by in fig \ref{OC}-a.\newline
On the other hand, by using the Hopf fibration $\mathbb{S}^{3}\sim \mathbb{S}%
^{1}\times \mathbb{S}^{2}$, which in Lie groups language corresponds to 
\begin{equation}
SU\left( 2\right) =U\left( 1\right) \times \frac{SU\left( 2\right) }{U\left(
1\right) }
\end{equation}
and using as well the toric fibration of the complex line $\mathbb{C}=%
\mathbb{R}^{+}\times \mathbb{S}^{1}$, one may also think about the conifold
as a complex 3d- toric manifold with the fibration 
\begin{equation}
\begin{tabular}{lll}
$\mathbb{C}\times \mathbb{S}^{2}\times \mathbb{S}^{2}$ & $\sim$ & $\mathbb{C}%
\times CP^{1}\times CP^{1}$%
\end{tabular}%
\end{equation}
The defining equations of this complex threefold, whose toric graph is given
by fig \ref{OC}-b, can be obtained from (\ref{2A}) by using the variable
change 
\begin{equation}
\begin{tabular}{lll}
$U_{1}=x_{1}+ix_{2}$ & , & $U_{2}=x_{3}+ix_{4}$ \\ 
$V_{1}=y_{1}+iy_{2}$ & , & $V_{2}=y_{3}+iy_{4}$%
\end{tabular}%
\end{equation}
to put it into the form%
\begin{equation}
\begin{tabular}{ll}
$\left \vert U_{1}\right \vert ^{2}+\left \vert U_{2}\right \vert ^{2}-\left
\vert V_{1}\right \vert ^{2}-\left \vert V_{2}\right \vert ^{2}$ & $=0 $ \\ 
$U_{1}\bar{V}_{1}+U_{2}\bar{V}_{2}+\bar{U}_{1}V_{1}+\bar{U}_{2}V_{2}$ & $=0$%
\end{tabular}
\label{U}
\end{equation}
These relations may be also obtained from eq(\ref{uv}) by setting 
\begin{equation}
\begin{tabular}{lll}
$u=U_{1}+iV_{1}$ & , & $z=U_{2}-iV_{2}$ \\ 
$v=\bar{U}_{1}+i\bar{V}_{1}$ & , & $w=\bar{U}_{2}-i\bar{V}_{2}$%
\end{tabular}%
\end{equation}
Notice that eqs(\ref{U}) are invariant under the $U\left( 1\right) $ gauge
symmetry%
\begin{equation}
\begin{tabular}{lll}
$U_{1}^{\prime}=e^{i\varphi}U_{1}$ & , & $V_{1}^{\prime}=e^{i\varphi}V_{1}$
\\ 
$U_{2}^{\prime}=e^{-i\varphi}U_{2}$ & , & $V_{2}^{\prime}=e^{-i\varphi}V_{2}$%
\end{tabular}%
\end{equation}
which is associated with the $\mathbb{S}^{1}$ fiber in the the Hopf
fibration of $\mathbb{S}^{3}$. Notice also that the compact part is given by
the relation $\left \vert U_{1}\right \vert ^{2}+\left \vert
U_{2}\right
\vert ^{2}=0$; and the small resolution of the conifold
singularity amounts to substitute it by%
\begin{equation}
\left \vert U_{1}\right \vert ^{2}+\left \vert U_{2}\right \vert
^{2}=\xi,\qquad U_{1}^{\prime}\equiv e^{i\varphi}U_{1},\qquad
U_{2}^{\prime}\equiv e^{-i\varphi}U_{2}  \label{k}
\end{equation}
where the positive number $\xi$ stands for the Kahler parameter of the
2-sphere $\mathbb{S}^{2}\sim \mathbb{S}^{3}/\mathbb{S}^{1}$. 
Under this resolution, eqs(\ref{U}) get promoted to%
\begin{equation}
\begin{tabular}{ll}
$\left \vert U_{1}\right \vert ^{2}+\left \vert U_{2}\right \vert ^{2}-\left
\vert V_{1}\right \vert ^{2}-\left \vert V_{2}\right \vert ^{2}$ & $=\xi$ \\ 
$U_{1}\bar{V}_{1}+U_{2}\bar{V}_{2}+\bar{U}_{1}V_{1}+\bar{U}_{2}V_{2}$ & $=0$%
\end{tabular}
\label{V}
\end{equation}
With these tools at hand we turn to study the zeros of the KW fermions.

\subsubsection{the $\mathcal{F}_{4}$ term (\protect \ref{F})}

Here, we give the explicit relation between the zeros of the $\mathcal{F}%
_{4} $ term (\ref{F}) and the small resolution of the conifold singularity.\
To that purpose, notice first that for $\alpha=0,$ we have%
\begin{equation}
\mathcal{F}_{4}\left( \alpha \right) |_{\alpha=0}=\sum_{l=1}^{4}\left(
1-\cos p_{l}\right)
\end{equation}
The typical terms $\left( 1-\cos p_{l}\right) $ can be seen as $\left(
1-z_{l}\right) $ with $z_{l}=\cos p_{l}$ giving the altitude of the
shrinking circle $\mathsf{x}_{l}^{2}+\mathsf{y}_{l}^{2}=\sin^{2}p_{l}$ of
the real 2-sphere parameterized by the spherical coordinates,%
\begin{equation}
\mathsf{x}_{l}=\rho_{l}\cos \varphi_{l},\qquad \mathsf{y}_{l}=\rho_{l}\sin
\varphi_{l},\qquad \mathsf{z}_{l}=\cos p_{l},
\end{equation}
with $\rho_{l}=\sin p_{l}$ and%
\begin{equation}
\mathsf{x}_{l}^{2}+\mathsf{y}_{l}^{2}+\mathsf{z}_{l}^{2}=1.
\end{equation}
Using the standard identities 
\begin{equation}
\begin{tabular}{lll}
$\sin p=2\cos \frac{p}{2}\sin \frac{p}{2}$ & , & $1-\cos p=2\sin^{2}\frac{p}{%
2}$%
\end{tabular}%
\end{equation}
we can rewrite the Dirac operator for the KW fermions like%
\begin{equation}
D_{KW}=2i\sum_{l=1}^{3}\gamma_{l}\cos \frac{p_{l}}{2}\sin \frac{p_{l}}{2}+%
\frac{2i}{\sin \alpha}\gamma_{4}\mathcal{F}_{4}
\end{equation}
with%
\begin{equation}
\mathcal{F}_{4}\left( p\right) =\sum_{l=1}^{4}\left( \sin \frac{p_{l}}{2}%
\right) ^{2}-\left( \sin \frac{\alpha}{2}\right) ^{2}
\end{equation}
Setting 
\begin{equation}
\begin{tabular}{lll}
$X_{l}=\cos \frac{p_{l}}{2}$ & , & $Y_{l}=\sin \frac{p_{l}}{2}$ \\ 
$A=\cos \frac{\alpha}{2}$ & , & $B=\sin \frac{\alpha}{2}$%
\end{tabular}
\label{AB}
\end{equation}
satisfying the constraint relation%
\begin{equation}
\begin{tabular}{lll}
$X_{l}^{2}+Y_{l}^{2}=1$ & , & $A^{2}+B^{2}=1$%
\end{tabular}
\label{CS}
\end{equation}
we can rewrite the $D_{KW}$ operator like 
\begin{equation}
D_{KW}=2i\sum_{l=1}^{3}\gamma_{l}X_{l}Y_{l}+\frac{2i}{\sin \alpha}\gamma _{4}%
\mathcal{F}_{4}  \label{K}
\end{equation}
with the $\mathcal{F}_{4}$ term along the $\gamma_{4}$ direction as follows 
\begin{equation}
\mathcal{F}_{4}\left( Y,B\right) =\left( Y_{1}\right) ^{2}+\left(
Y_{2}\right) ^{2}+\left( Y_{3}\right) ^{2}+\left( Y_{4}\right) ^{2}-B^{2}
\label{4F}
\end{equation}
Clearly, in the real coordinate frame $\left( Y_{1},Y_{2},Y_{3},Y_{4}\right) 
$, the zeros of the $\mathcal{F}_{4}$ term describe a real 3-sphere $\mathbb{%
S}^{3}$ of radius $\left \vert B\right \vert $. So the the limit $%
B\rightarrow0$, which by using (\ref{AB}) corresponds to the limit $\alpha
\rightarrow0$, describes precisely the vanishing of the volume of a 3-sphere;%
\begin{equation}
\mathcal{F}_{4}\left( Y,B\right) |_{B=0}=0\qquad \Leftrightarrow \qquad
\left( Y_{1}\right) ^{2}+\left( Y_{2}\right) ^{2}+\left( Y_{3}\right)
^{2}+\left( Y_{4}\right) ^{2}=0
\end{equation}
This shows that the of $\mathcal{F}_{4}\left( Y,B\right) |_{B\neq0}$ is
related to the resolution of a 3-dimensional conifold singularity of fig \ref%
{OC}-b.

\subsection{$\mathcal{F}_{4}$- term and the complexified $D_{KW}$ operator}

Following the same reasoning we have done for naive fermions, one may think
about the the antihermitian operator $D_{KW}$ as given by 
\begin{equation}
D_{KW}=\frac{D-D^{+}}{2}
\end{equation}
with%
\begin{equation}
D=\sum_{l=1}^{3}\gamma_{l}z_{l}^{2}+\frac{1}{c^{2}}\gamma_{4}\left(
z_{1}^{2}+z_{2}^{2}+z_{3}^{2}+z_{4}^{2}-c^{2}\right)  \label{W}
\end{equation}
where the $z_{l}$'s are complex variables and $c$ is a priori a complex
modulus. Clearly $D$ contains (\ref{K}) and is singular for $c=0$ as in the
case of $D_{KW}$. Writing this complex operator $D$ like, 
\begin{equation}
D=\sum_{l=1}^{3}\gamma_{l}\mathcal{L}_{l}+\frac{1}{c^{2}}\gamma_{4}\mathcal{L%
}_{4}
\end{equation}
with%
\begin{equation}
\begin{tabular}{lll}
$\mathcal{L}_{l}$ & $=\frac{1}{2}\left \{ \gamma_{l},D\right \} $ & $%
=z_{l}^{2}$ \\ 
$\mathcal{L}_{4}$ & $=\frac{c^{2}}{2}\left \{ \gamma_{4},D\right \} $ & $%
=\left( z_{1}^{2}+z_{2}^{2}+z_{3}^{2}+z_{4}^{2}-c^{2}\right) $%
\end{tabular}%
\end{equation}
Then substituting 
\begin{equation}
z_{l}=x_{l}+iy_{l},\qquad c=\xi+i\zeta
\end{equation}
back into (\ref{W}), we get for the first 3 terms $\mathcal{L}_{1}$, $%
\mathcal{L}_{2}$, $\mathcal{L}_{3}$ the generic expression%
\begin{equation}
\mathcal{L}_{l}=\left( x_{l}^{2}-y_{l}^{2}\right) +2ix_{l}y_{l}
\end{equation}
and for the 4-th the following 
\begin{equation}
\mathcal{L}_{4}=\frac{\xi-i\zeta}{\xi^{2}+\zeta^{2}}\left( \dsum
\limits_{l=1}^{4}\left( x_{l}^{2}-y_{l}^{2}\right) -\left(
\xi^{2}-\zeta^{2}\right) \right) +2i\frac{\xi-i\zeta}{\xi^{2}+\zeta^{2}}%
\left( \dsum \limits_{l=1}^{4}x_{l}y_{l}-\xi \zeta \right)
\end{equation}
In the particular case $\xi=0$, the term $\mathcal{L}_{4}$ reduces to%
\begin{equation}
\mathcal{L}_{4}=\frac{i}{\zeta}\left( \dsum \limits_{l=1}^{4}\left(
x_{l}^{2}-y_{l}^{2}\right) +\zeta^{2}\right) +\frac{2}{\zeta}\dsum
\limits_{l=1}^{4}x_{l}y_{l}
\end{equation}
By setting%
\begin{equation}
\begin{tabular}{lll}
$u_{1}=x_{1}+ix_{3}$ & , & $u_{2}=x_{2}+ix_{4}$ \\ 
$v_{1}=y_{1}+iy_{3}$ & , & $v_{2}=y_{2}+iy_{4}$%
\end{tabular}%
\end{equation}
we also have%
\begin{align}
\mathcal{L}_{4} & =\frac{-i}{\zeta}\left( \left \vert v_{1}\right \vert
^{2}+\left \vert v_{1}\right \vert ^{2}-\left \vert u_{1}\right \vert
^{2}-\left \vert u_{1}\right \vert ^{2}-\zeta^{2}\right)  \notag \\
& -\frac{2}{\zeta}\left( u_{1}\bar{v}_{1}+u_{2}\bar{v}_{2}+\bar{u}_{1}v_{1}+%
\bar{u}_{2}v_{2}\right)
\end{align}
The zeros of $\mathcal{L}_{4}$ are obtained by solving the two following
relations given by the real and imaginary parts%
\begin{equation}
\begin{tabular}{lll}
$\left \vert v_{1}\right \vert ^{2}+\left \vert v_{1}\right \vert ^{2}-\left
\vert u_{1}\right \vert ^{2}-\left \vert u_{1}\right \vert ^{2}$ & $%
=\zeta^{2}$ &  \\ 
$u_{1}\bar{v}_{1}+u_{2}\bar{v}_{2}+\bar{u}_{1}v_{1}+\bar{u}_{2}v_{2}$ & $=0$
& 
\end{tabular}%
\end{equation}
But these relations are nothing but those describing the resolved conifold (%
\ref{V}). The compact part of this manifold is given by $\left \vert
v_{1}\right \vert ^{2}+\left \vert v_{1}\right \vert ^{2}=\zeta^{2}$ as
given by \ref{OC}-(c); the non compact part is parameterized by the $u_{l}$%
's.

\section{Conclusion and comment}

Motivated by Creutz point splitting method of refs \textrm{\cite{A1,A2}}, we
have learnt in this study two basic things regarding naive and
Karsten-Wilczek fermions of lattice QCD. Concerning naive fermions, we have
shown that the zeros of the Dirac operator $D_{naive}$ have a geometric
interpretation in terms of \emph{toric singularities }of some complex Kahler
manifolds that have been explicitly constructed here. The Brillouin zone
moded by the $\mathbb{Z}_{2}$ antisymmetry (\ref{Z2}) of the Dirac operator $%
D_{naive}$ turns out to be exactly the base of the toric fibration of a
toric Kahler manifold $\mathcal{K}$. For instance, in the case of naive
fermions of QCD$_{2}$, the corresponding toric manifold is precisely the
complex surface 
\begin{equation}
CP^{1}\times CP^{1}\text{ }\sim \text{ }\mathbb{S}^{2}\times \mathbb{S}^{2}
\end{equation}
with toric graph given by the fig \ref{OK}. Moreover seen that the
homogeneous coordinates $\left( z,\zeta \right) $ of each $CP^{1}$ form an $%
SU\left( 2\right) $ doublet, it follows that the 4 zeros of naive fermions
of QCD$_{2}$ live at the poles of $\mathbb{S}^{2}\times \mathbb{S}^{2}$ and
carry quantum charges of the bi-spinor representation 
\begin{equation}
(\frac{1}{2},\frac{1}{2})\text{ \  \ of \  \  \ }SU\left( 2\right) \times
SU\left( 2\right) \sim SO\left( 4\right)
\end{equation}
This result extends straightforwardly to higher dimensions. In QCD$_{4}$,
the Dirac operator of the naive fermions in the reciprocal space 
\begin{equation}
D_{naive}=i\gamma_{1}\sin p_{1}+i\gamma_{2}\sin p_{2}+i\gamma_{3}\sin
p_{3}+i\gamma_{4}\sin p_{4}
\end{equation}
has $2^{4}=16$ zeros located at $\mathbf{P}_{\left(
m_{1},m_{2},m_{3},m_{4}\right) }=$ $(P_{1}^{\left( m_{1}\right)
},...,P_{4}^{\left( m_{1}\right) })$ with%
\begin{equation}
\begin{tabular}{lll}
$P_{1}^{\left( m_{1}\right) }=m_{1}\pi,$ & $P_{2}^{\left( m_{2}\right)
}=m_{2}\pi,$ &  \\ 
$P_{3}^{\left( m_{3}\right) }=m_{3}\pi,$ & $P_{4}^{\left( m_{4}\right)
}=m_{4}\pi,$ & 
\end{tabular}
,\qquad m_{l}=0,1  \label{ze}
\end{equation}
Because of periodicity of $D_{naive}$ in the wave phases $p_{l}$ and because
of the $\mathbb{Z}_{2}$ anti-symmetry changing the signs of these phases,
the zeros of the 4 $\sin p_{l}$'s, which are also fix points of $\mathbb{Z}%
_{2}$, should be viewed as the vertices of 4d- unit cell in the reciprocal
space. This unit cell has 3-dimensional divisors (cubes), 2-dimensional
faces (squares), 1-dimensional edges (segments) and vertices (points). On
the 3d-divisors one of the 4 $\sin p_{l}$ vanish while on the 2d- faces two
of them vanish. On the edges of the unit cell, 3 of the 4 $\sin p_{l}$
vanish and at the vertices the 4 $\sin p_{l}$'s vanish in complete agreement
with toric singularities of the toric four-fold%
\begin{equation}
CP^{1}\times CP^{1}\times CP^{1}\times CP^{1}  \label{4}
\end{equation}
Moreover, the expansions of the Dirac operator near the 16 zeros read in a
condensed manner like 
\begin{equation}
D_{\mathbf{m}}\simeq \gamma_{\left( m_{1}\right) }^{1}q_{1}+\gamma_{\left(
m_{2}\right) }^{2}q_{2}+\gamma_{\left( m_{3}\right)
}^{3}q_{3}+\gamma_{\left( m_{4}\right) }^{4}q_{4}+O\left( q^{2}\right)
\end{equation}
where $\mathbf{m}$ stand for the 4-dimensional integral vector $\left(
m_{1},m_{2},m_{3},m_{4}\right) $ with $m_{l}=0,1;$ and where the gamma
matrices $\gamma_{\left( m_{i}\right) }^{i}$ are related to the standard
ones $\gamma^{l}$ as follows 
\begin{equation}
\gamma_{\left( m_{i}\right) }^{i}=\left( -\right) ^{m_{i}}\gamma^{i}
\end{equation}
The wave functions $\Psi_{\mathbf{m}}=$ $\psi \left( q_{\mathbf{m}}\right) $
near the 2$^{4}$ zero modes (\ref{ze}) with 
\begin{equation}
q_{\mathbf{m}}=(p_{1}-P_{1}^{\left( m_{1}\right) },p_{2}-P_{2}^{\left(
m_{2}\right) },p_{3}-P_{3}^{\left( m_{3}\right) },p_{4}-P_{4}^{\left(
m_{1}\right) })
\end{equation}
depend on the values of the integral vector $\mathbf{m}$ capturing the $%
\left[ SU\left( 2\right) \right] ^{4}$ symmetry of the four-fold (\ref{4}).
So the waves $\Psi_{\mathbf{m}}$ carry quantum charges of the fundamental
representation 
\begin{equation}
(\frac{1}{2},\frac{1}{2},\frac{1}{2},\frac{1}{2})
\end{equation}
of\ the group $\left[ SU\left( 2\right) \right] ^{4}$.\  \newline
Concerning the Karsten-Wilcek fermions of lattice QCD$_{4}$, we have learnt
that the two zero modes of the Dirac operator $D_{KW}$ (\ref{3}) are fix
points of the $\mathbb{Z}_{2}$ symmetry (\ref{z}). Like for naive fermions,
these zeros for an $SU\left( 2\right) $ doublet and have as well a geometric
interpretation in terms of the small resolution of the conifold.

\section{Appendix: Features of $\mathbb{S}^{2}\sim CP^{1}$}

In this appendix, we collect some useful relations on the real 2-sphere $%
\mathbb{S}^{2}$. Because of special features, this compact 2-dimensional
surface can be approached in various ways; three of them were briefly
described in subsection 4.1, eqs(\ref{x}) to (\ref{y}); they are:%
\begin{equation}
\begin{tabular}{llll}
$\left( i\right) $ & $:$ & $\mathbb{S}^{2}$ & $\subset \mathbb{R}^{3}$ \\ 
$\left( ii\right) $ & $:$ & $\mathbb{S}^{2}$ & $\simeq SU\left( 2\right)
/U\left( 1\right) $ \\ 
$\left( iii\right) $ & $:$ & $\mathbb{S}^{2}$ & $\simeq CP^{1}$%
\end{tabular}%
\end{equation}
Our interest into the real 2-sphere; in particular into its Kahler structure
and toric representation, is because the north and south poles of $\mathbb{S}%
^{2}$ host the Creutz flavors of naive fermions considered in this paper. 
\newline
First, notice that the power of the properties of $\mathbb{S}^{2}$ comes
essentially from its link with the complex projective line and the $SU\left(
2\right) $ Lie group. To any complex number $z=x+iy\in \mathbb{C}$,
represented as $\left( x,y\right) $ in the plane $\mathbb{R}^{2}$, one can
associate a point $\left( u,v,w\right) $ on the unit 2-sphere 
\begin{equation}
\mathbb{S}^{2}=\left \{ \left( u,v,w\right) \in \mathbb{R}^{3}\text{ \ }|%
\text{ \  \ }u^{2}+v^{2}+w^{2}=1\right \}
\end{equation}
This mapping is known as the stereographic projection from the unit sphere $%
\mathbb{S}^{2}$ minus the point $N=\left( 0,0,1\right) $ (north\footnote{%
one may also consider the stereographic projection using the south pole $%
\left( 0,0,-1\right) $.} pole) onto the plane $w=0$, which we identify with
the complex plane with coordinate $z=x+iy$.

\begin{figure}[ptbh]
\begin{center}
\hspace{0cm} \includegraphics[width=8cm]{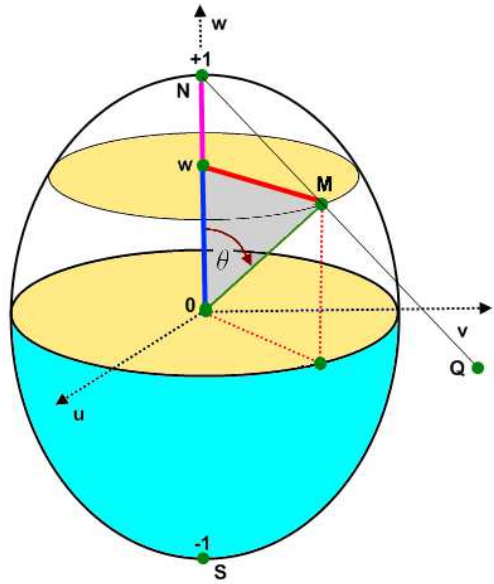}
\end{center}
\par
\vspace{0cm}
\caption{stereographic projection of real 2-sphere using north pole. Recall
that the poles of the sphere host Creutz fermions; the coefficient $\sin
p_{l}$ of the Dirac operator of naive fermions are interpreted in terms of
the radius $\sin \protect \theta$ of the parallel circle which shrinks at the
poles.}
\label{PS}
\end{figure}
To get the explicit relations between the $\left( u,v,w\right) $ variables
and the planar $\left( x,y\right) $, one projects the point $P=\left(
u,v,w\right) $ $\in$ $\mathbb{S}^{2}-\left \{ N\right \} $ on to the plane $%
w=0$. Straightforward calculations lead to%
\begin{equation}
\begin{tabular}{lllll}
$u=\frac{2x}{1+\left \vert z\right \vert ^{2}}$ & , & $v=\frac{2y}{1+\left
\vert z\right \vert ^{2}}$ & , & $w=\frac{\left \vert z\right \vert ^{2}-1}{%
1+\left \vert z\right \vert ^{2}}$%
\end{tabular}%
\end{equation}
with inverse relations as%
\begin{equation}
\begin{tabular}{lll}
$x=\frac{u}{1-w}$ & , & $y=\frac{v}{1-w}$%
\end{tabular}%
\end{equation}
In the spherical coordinates $\left( \theta,\varphi \right) $ with zenith $%
\theta$ and azimuth $\phi$ variables as $0\leq \theta \leq \pi$, $0\leq
\varphi \leq2\pi$; we can combine the $u$ and $v$ coordinates into a complex
one like 
\begin{equation}
u+iv=\sin \theta e^{i\varphi},\qquad w=\cos \theta
\end{equation}
This leads to $\left \vert u+iv\right \vert =\sin \theta$ and shows that the
radius $\rho=\sin \theta$ of the parallel circle in $\mathbb{S}^{2}$
vanishes at north $\theta=0$ and south $\theta=\pi$ poles. By setting 
\begin{equation}
r=\frac{\sin \theta}{1-\cos \theta}=\frac{\cos \frac{\theta}{2}}{\sin \frac {%
\theta}{2}}  \label{st}
\end{equation}
we also have%
\begin{equation}
x+iy=re^{i\varphi},\qquad r=\cot \frac{\theta}{2}  \label{ts}
\end{equation}
showing that the singularity at the south pole $\theta=\pi$ is
stereographically mapped to the origin of the complex plane; and the
singularity at north $\theta=0$ is mapped to infinity. This property should
be compared with the two zeros of the complex ratio $\frac{z}{\zeta}$
describing the complex projective line parameterized by the homogeneous
coordinates $\left( z,\zeta \right) $; see also eqs(\ref{D}) and \ref{zz})
to make contact with zero modes of the Dirac operator of naive fermions and 
\emph{Creutz} flavors.\newline
Seen as a Khaler manifold, the metric of the complex projective line follows
as a particular case of the Fubini-Study metric of complex n-dimensional
projective spaces $CP^{n}$; it reads in the local coordinate patch (chart) $%
\zeta=1$ as follow%
\begin{equation}
ds^{2}=\frac{4}{\left( 1+\left \vert z\right \vert ^{2}\right) ^{2}}dzd\bar{z%
}
\end{equation}
Automorphisms of the 2-sphere using the (non homogenous) affine coordinate $%
z $ is given by the mobius transformations 
\begin{equation}
z^{\prime}=\frac{az+b}{cz+d}
\end{equation}
with $a,b,c,d$ four complex numbers constrained as $ad-bc\neq0$. By using
the homogeneous complex coordinates $\left( z,\zeta \right) $ , this
transformation reads also like%
\begin{equation}
\left( 
\begin{array}{c}
z^{\prime} \\ 
\zeta^{\prime}%
\end{array}
\right) =\left( 
\begin{array}{cc}
a & b \\ 
c & d%
\end{array}
\right) \left( 
\begin{array}{c}
z \\ 
\zeta%
\end{array}
\right)
\end{equation}
and shows that $\left( z,\zeta \right) $ transform indeed as an $SU\left(
2\right) $ doublet.

\begin{acknowledgement}
This work is supported by CNRST under contract URAC09.
\end{acknowledgement}


\begin{thebibliography}{99}
\bibitem{A1} M. Creutz, Minimal doubling and point splitting, PoS: Lattice
2010, 078 (2010) [arXiv:1009.3154],

\bibitem{A2} Michael Creutz, Taro Kimura, Tatsuhiro Misumi, \emph{Index
Theorem and Overlap Formalism with Naive and Minimally Doubled Fermions},
JHEP 1012:041,2010, arXiv:1011.0761

\bibitem{B1} Michael Creutz, Taro Kimura, Tatsuhiro Misumi, Aoki Phases in
the Lattice Gross-Neveu Model with Flavored Mass terms,
Phys.Rev.D83:094506,2011, arXiv:1101.4239,

\bibitem{2B} Michael Creutz, \emph{Confinement, chiral symmetry, and the
lattice}, Acta Physica Slovaca 61, No.1, 1-127 (2011), arXiv:1103.3304,

\bibitem{B2} P. F. Bedaque, M. I. Buchoff, B. C. Tiburzi and A. Walker-Loud,
Phys. Lett. B 662, 449 (2008) [arXiv:0801.3361],

\bibitem{B3} P. F. Bedaque, M. I. Buchoff, B. C. Tiburzi and A. Walker-Loud,
Phys. Rev. D 78, 017502 (2008) [arXiv:0804.1145]

\bibitem{B4} D. H. Adams, Phys. Rev. Lett. 104, 141602 (2010)
[arXiv:0912.2850],

\bibitem{B5} D. H. Adams, Phys. Lett. B 699, 394 (2011), (2010)
[arXiv:1008.2833],

\bibitem{B6} E.H Saidi et al, \emph{Topological\ Aspects of Fermions on
Hyperdiamond, }LPHE preprint, October 2011,

\bibitem{B7} L.B Drissi, H. Mhamdi, E.H Saidi, \emph{Anomalous Quantum Hall
Effect of 4D Graphene in Background Fields,} JHEP 1110:026,2011,
arXiv:1106.5578,

\bibitem{AB1} Michael Creutz, JHEP 04 (2008) 017,[arXiv:0712.1201],

\bibitem{AB2} A.Borici, Phys. Rev. D78 (2008) 074504, [arXiv:0712.4401]

\bibitem{AB3} Michael Creutz, Tatsuhiro Misumi, \emph{Classification of
Minimally Doubled Fermions}, Phys.Rev.D82:074502,2010, arXiv:1007.3328,

\bibitem{AB4} S. Capitani, M. Creutz, J. Weber, H.Wittig, JHEP
1009:027,2010, arXiv:1006.2009,

\bibitem{AB5} L.B Drissi, E.H Saidi, M. Bousmina, \emph{Electronic
Properties and Hidden Symmetries of Graphene}, Nucl.Phys.B829:523-533,2010,
arXiv:1008.4470,

\bibitem{AB6} L.B Drissi, E.H Saidi, M. Bousmina, \emph{Graphene, Lattice
QFT and Symmetries}, J. Math. Phys. 52:022306,2011, arXiv:1101.1061

\bibitem{C1} S. Katz, P. Mayr, C. Vafa, Adv.Theor.Math.Phys. 1 (1998)
53-114, hep-th/9706110,

\bibitem{C2} A. Belhaj, E.H.Saidi, \emph{On HyperKahler Singularities},
Mod.Phys.Lett. A15 (2000) 1767-1780, arXiv:hep-th/0007143,

\bibitem{D0} N.C. Leung and C. Vafa; Adv .Theo . Math. Phys 2(1998) 91,
hep-th/9711013,

\bibitem{D} P. Candelas and X. de la Ossa, \emph{\textquotedblleft Comments
on Conifolds,}\textquotedblright \ Nucl. Phys. B342 (1990), 246,

\bibitem{1D} E. Witten, \emph{Branes and the dynamics of QCD, }Nucl. Phys.
B507 (1997) 658, hep-th/9706109

\bibitem{D1} Harald Nieder, Yaron Oz, \emph{Supergravity and D-branes
Wrapping Supersymmetric 3-Cycles}, JHEP 0103:008,2001, arXiv:hep-th/0011288,

\bibitem{D2} Robbert Dijkgraaf, Lotte Hollands, Piotr Sulkowski, Cumrun Vafa,%
\emph{\ Supersymmetric Gauge Theories, Intersecting Branes and Free
Fermions, }JHEP 0802:106,2008, arXiv:0709.4446,

\bibitem{D3} Martijn Wijnholt, F-Theory, GUTs and Chiral Matter,
Fortschritte Der Physik-progress of Physics - FORTSCHR PHYS , vol. 58, no.
7-9, pp. 846-854, 2010, arXiv:0809.3878

\bibitem{D4} Chris Beasley, Jonathan J. Heckman, Cumrun Vafa, GUTs and
Exceptional Branes in F-theory - I, JHEP 0901:058,2009, arXiv:0802.3391

\bibitem{D5} Lalla Btissam Drissi, Leila Medari, El Hassan Saidi, \emph{%
Quiver Gauge Models in F-Theory on Local Tetrahedron}, arXiv:0908.0471.

\bibitem{E2} Mohamed Bennai, El Hassan Saidi, \emph{Toric Varieties with NC
Toric Actions: NC Type IIA Geometry}, Nucl.Phys.B677:587-613,2004,
arXiv:hep-th/0312200,

\bibitem{E3} Mohamed Bennai, El Hassan Saidi, \emph{NC Calabi-Yau Manifolds
in Toric Varieties with NC Torus fibration}, Phys.Lett.B550:108-116,2002,
hep-th/0210073,

\bibitem{E1} El Hassan Saidi, \emph{F-theory on tetrahedron}, Frontiers in
Science and Engineering, An International Journal Edited by Hassan II
Academy of Science and Technology, Vol 1 (2011) pages 1-24,

\bibitem{F1} L. H. Karsten, Phys. Lett. B 104, 315 (1981),

\bibitem{F2} F.Wilczek, Phys. Rev. Lett. 59, 2397 (1987),

\bibitem{G1} Igor R. Klebanov, Matthew J. Strassler, \emph{Supergravity and
a Confining Gauge Theory: Duality Cascade}, HEP 0008:052,2000,
arXiv:hep-th/0007191,

\bibitem{G2} Rachid Ahl Laamara, Lalla Btissam Drissi, El Hassan Saidi, 
\emph{D-string fluid in conifold: I. Topological gauge model},
Nucl.Phys.B743:333-353,2006, arXiv:hep-th/0604001,

\bibitem{G3} Rachid Ahl Laamara, Lalla Btissam Drissi, El Hassan Saidi, 
\emph{D-string fluid in conifold: II. Matrix model for D-droplets on S}$^{3}$%
\emph{\ and S}$^{2}$, Nucl.Phys. B749 (2006) 206-224, arXiv:hep-th/0605209,

\bibitem{G4} El Hassan Saidi, \emph{Topological SL(2) Gauge Theory on
Conifold}, African Journal Of Mathematical Physics Vol 5 (2007)57-77,
arXiv:hep-th/0601020

\bibitem{H1} L.B Drissi, E.H Saidi, M. Bousmina, \emph{Four Dimensional
Graphene}, Phys.Rev.D84:014504,2011, arXiv:1106.5222,

\bibitem{H2} Lalla Btissam Drissi, El Hassan Saidi, \emph{On Dirac Zero
Modes in Hyperdiamond Model}, Phys.Rev.D84:014509,2011, arXiv:1103.1316,

\bibitem{I1} El Hassan Saidi, \emph{Tetrahedron in F-theory Compactification}%
, arXiv:0907.2655,

\bibitem{J1} Malika Ait Benhaddou, El Hassan Saidi, \emph{Explicit Analysis
of Kahler Deformations in 4D N=1 Supersymmetric Quiver Theories, }Physics
Letters B575(2003)100-110, arXiv:hep-th/0307103

\bibitem{J2} T. Eguchi, P.B Gilkey, A.J Hanson, \emph{Gravitation Gauge
Theories And Differential Geometry}, Physics Reports 66 No 6 pp 213 393
(1980)
\end{thebibliography}
\end{document}